\documentclass[11pt,a4paper]{article}
\usepackage{jheppub}
\usepackage{slashed}
\usepackage{arydshln}
\usepackage{mathrsfs}
\usepackage{dsfont}
\usepackage{braket}
\usepackage{bbm}

\newcommand\cl{{\cal L}}
\newcommand*\DAlambert{\mathop{}\!\mathbin\Box}

\newcommand{\g}{\gamma}

\newcommand\bbone{\ensuremath{\mathbbm{1}}}

\newcommand\beal{\begin{align}}

\renewcommand{\theenumii}{\alph{enumii}}

\newcommand{\eq}[1]{\begin{equation}#1\end{equation}}
\newcommand{\al}[1]{\begin{align}#1\end{align}}

\newcommand{\spl}[1]{\begin{split}#1\end{split}}

\title{3d \boldmath$\mathcal{N} = 1$ Chern-Simons-matter theory\\ and localization}
\author[a]{Dimitrios Tsimpis,}
\author[b]{Yaodong Zhu}
\affiliation[a]{Universit\'{e} de Lyon\\
UMR 5822, CNRS/IN2P3, Institut de Physique Nucl\'{e}aire de Lyon\\
4 rue Enrico Fermi, 69622 Villeurbanne Cedex,  France}
\affiliation[b]{George and Cynthia Woods Mitchell Institute for Fundamental Physics and Astronomy,\\
Texas A\&M University, College Station, TX 77843, USA}
\emailAdd{tsimpis@ipnl.in2p3.fr}
\emailAdd{yaodongmatt@physics.tamu.edu}
\abstract{
We consider the most general, classically-conformal, three-dimensional $\mathcal{N}=1$ Chern-Simons-matter theory  with global symmetry $Sp(2)$ and gauge group $U(N)\times U(N)$. We show that 
the Lagrangian in the on-shell formulation of the theory admits one more free parameter  as compared to the theory formulated in off-shell $\mathcal{N}=1$  superspace. 
The theory on $T^3$ can be formally localized. 
We partially carry out the localization procedure for the theory on $T^3$ with periodic boundary conditions. In particular we show that restricting to the saddle points with vanishing gauge connection gives a trivial contribution to the partition function, i.e.~the bosonic and fermionic contributions exactly cancel each other.
}

\begin{document}
\maketitle
\flushbottom

\section{Introduction}

Using localization, several exact results have by now been obtained for supersymmetric gauge theories, such as  the computation of indices, partition functions and Wilson loops, among others. In many cases these exact computations have provided us with checks of non-trivial dualities, including AdS/CFT. 

In the present paper we apply the localization procedure to the most general, classically-conformal, three-dimensional $\mathcal{N}=1$ Chern-Simons-matter theory  with global symmetry $Sp(2)$ and gauge group $U(N)\times U(N)$. 
Previously, localization had mainly been used to study theories on curved spacetimes with non-trivial R-symmetry \cite{Pestun:2007rz,Kapustin:2009kz,Hama:2010av,Hama:2011ea,Tanaka:2013dca,
Alday:2013lba,Imamura:2011wg,Kallen:2011ny,Ohta:2012ev,Nian:2013qwa,Imbimbo:2014pla}. We show that the $\mathcal{N}=1$ theory on a flat three-dimensional torus $T^3$ can also be formally localized. 

The $\mathcal{N}=1$  theory we consider here is not in general superconformal on the quantum level, except for special points in its moduli space where supersymmetry may be enhanced. In particular the ABJM model \cite{Aharony:2008ug} is one such special point where 
supersymmetry is enchanced to $\mathcal{N}=6$. By considering a classically-conformal $\mathcal{N}=1$ theory with unequal Chern-Simons (CS) levels  which is in a certain sense a small deformation of the ABJM model \cite{Aharony:2008ug}, it was argued in \cite{Gaiotto:2009mv}  that the theory flows to an RG fixed point in the infrared. These CFT's were then conjectured in \cite{Gaiotto:2009mv} to be dual to certain (massive) IIA supergavity solutions \cite{Tomasiello:2007eq,Koerber:2008rx} which fall within the general class of \cite{Lust:2004ig}.

The outline of the paper is as follows. In section 2 we give the on-shell formulation of the most general classically-conformal $\mathcal{N}=1$ $U(N)\times U(N)$ CS-matter theory with  $Sp(2)$ global symmetry. We then introduce auxiliary fields and formulate the theory off-shell, as required by the localization procedure. An interesting observation is that the Lagrangian in the on-shell formulation of the theory admits one more free parameter  as compared to the theory formulated in off-shell  $\mathcal{N}=1$ superspace.

In section 3 we formulate the theory on a curved manifold. One notable difference from the CS theories with $\mathcal{N}=2$ supersymmetry studied in \cite{Kapustin:2009kz} is that the requirement of localization excludes  positive-curvature manifolds such as $S^3$. Formulating the theory on $T^3$ or the hyperbolic three-dimensional space $H_3$ preserves superconformal symmetry at the classical level. In this paper we shall focus on the theory on  $T^3$.\footnote{Demanding that the manifold should be compact, in order to ensure that the partition function is well-defined, leads us to exclude $H_3$. Compact quotients thereof may still preserve superconformal symmetry but we shall not examine this possibility here.}

We next carry out the localization procedure for the theory on $T^3$ with periodic boundary conditions. As an illustration of the formalism we compute the contributions to the partition function from the locus of saddle points with vanishing gauge connection. We show that restricting to this locus gives a trivial contribution to the partition function, i.e.~the bosonic and fermionic contributions exactly cancel each other. We conclude with a discussion of our results in section 4. 
Further technical details can be found in the appendices.

\vfill\break

\section{$\mathcal{N}=1$ Superconformal Chern-Simons-matter theory}

\subsection{On-Shell}

The general component form of the on-shell $\mathcal{N}=1$ classically-superconformal CS Lagrangian with $Spin(5)\simeq Sp(2)$ global symmetry and gauge group $U(N)\times U(N)$ is given in \cite{Ooguri:2008dk}:
\begin{equation}
\label{r20}
\cl=\cl_{CS}+\cl_{kin}+\cl_{4}+\cl_{6}~, 
\end{equation}
where $\cl_{CS}$ is the pure CS Lagrangian, $\cl_{kin}$ is the matter kinetic term, $\cl_{4}$ is the quartic interaction and $\cl_{6}$ is the sextic potential. More specifically,\footnote{We follow closely the notation of \cite{Ooguri:2008dk}, 
to which the reader is referred for more details; 
our spinor notation is explained in appendix A.}
\begin{equation}
\cl_{CS}=\frac{k_1}{2\pi}\varepsilon^{\mu\nu\rho}\mathrm{tr}\left\{
\frac{1}{2}A_{\mu}\partial_{\nu}A_{\rho}+\frac{i}{3}A_{\mu}A_{\nu}A_{\rho} \right\}
-\frac{k_2}{2\pi}\varepsilon^{\mu\nu\rho}\mathrm{tr}\left\{
\frac{1}{2}\hat{A}_{\mu}\partial_{\nu}\hat{A}_{\rho}
+\frac{i}{3}\hat{A}_{\mu}\hat{A}_{\nu}\hat{A}_{\rho}
  \right\}~,
\end{equation}
where the normalization above was chosen to facilitate the derivation of the 
superconformal invariance; ${A}_{\mu}$, $\hat{A}_{\mu}$ are gauge fields in the adjoint of $U(N)$.  
The matter kinetic terms read:
\begin{equation}
\cl_{kin}=\frac{1}{2\pi}\mathrm{tr}\left\{
-D^{\mu}X^AD_{\mu}X_A+i\tilde{\Psi}_A\gamma^{\mu}D_{\mu}\Psi^A
\right\}~,
\end{equation}
where $A=1,\dots,4$ is an $Sp(2)$ index; 
$X_A$ is in the bifundamental $(\bar{N},N)$ while $X^A$ is in the 
$(N,\bar{N})$, and similarly for $\Psi_A$, $\Psi^A$. 
The most general quartic interaction terms can be written in the form
$\cl_4=\cl_{4a}+\cl_{4b}+\cl_{4c}+\cl^\prime$, where:
\begin{equation}
\begin{split}
\cl_{4a}&=\frac{1}{2\pi}i\mathrm{tr}\{\bar{\alpha}_1\varepsilon^{ABCD}\tilde{\Psi}_AX_B\Psi_CX_D-\alpha_1\varepsilon_{ABCD}\tilde{\Psi}^AX^B\Psi^CX^D\} \\
\cl_{4b}&=\frac{1}{2\pi}i\mathrm{tr}\{\alpha_{2,1}\tilde{\Psi}^A\Psi_AX_BX^B-\alpha_{2,2}\tilde{\Psi}_A\Psi^AX^BX_B\} \\
\cl_{4c}&=\frac{1}{2\pi}2i\mathrm{tr}\{\alpha_{3,1}\tilde{\Psi}_A\Psi^BX^AX_B-\alpha_{3,2}\tilde{\Psi}^B\Psi_AX_BX^A\} \\
\cl^\prime&=\frac{1}{2\pi}\mathrm{tr}\{a_1\Omega^{AD}\Omega_{BC}\tilde{\Psi}_A\Psi^BX^CX_D+a_2\Omega_{AD}\Omega^{BC}\tilde{\Psi}^A\Psi_BX_CX^D \\
&~~~~~~~~~+a_3\Omega^{AC}\Omega^{BD}\tilde{\Psi}_AX_B\Psi_CX_D+\bar{a}_3\Omega_{AC}\Omega_{BD}\tilde{\Psi}^AX^B\Psi^CX^D \\
&~~~~~~~~~+a_4\Omega^{AB}\Omega^{CD}\tilde{\Psi}_AX_B\Psi_CX_D+\bar{a}_4\Omega_{AB}\Omega_{CD}\tilde{\Psi}^AX^B\Psi^CX^D\}~.
\end{split}
\end{equation}
The sextic potential consists of two terms $\cl_{6}=\cl_{pot}+\cl^{\prime \prime}$, where:
\begin{equation}
\begin{split}
\cl_{pot}=&\frac{1}{2\pi}\frac{1}{3}\mathrm{tr} \{\alpha_{4,1}X^AX_AX^BX_BX^CX_C+\alpha_{4,2}X_AX^AX_BX^BX_CX^C \\
&~~~~~~~~~+4\alpha_{4,3}X_AX^BX_CX^AX_BX^C-6\alpha_{4,4}X^AX_BX^BX_AX^CX_C\} \\
\cl^{\prime \prime}=&\frac{1}{2\pi}\Omega^{BC}\Omega_{DE}\mathrm{tr}\{
n X_BX^AX_CX^DX_AX^E\} \\
&+\frac{1}{2\pi}\Omega^{BC}\Omega_{DE}\mathrm{tr}\{mX_BX^AX_AX^DX_CX^E\}\\
&+\frac{1}{2\pi}\Omega_{BC}\Omega^{DE}\mathrm{tr}\{\bar{m}X^BX_AX^AX_DX^CX_E\}~.
\end{split}
\end{equation}
Here $\Omega_{AB}$ is the $Sp(2)$-invariant antisymmetric tensor, which satisfies $\Omega^{AB}\Omega_{AC}=\delta^B_C$. As shown in Appendix B, the theory is invariant under the following $\mathcal{N}=1$ Poincar\'e supersymmetry:
\begin{equation}
\begin{split}
\delta X_A=&i\Omega_{AB}\tilde{\epsilon}\Psi^B\\
\delta X^A=&i\Omega^{AB}\tilde{\epsilon}\Psi_B\\
\delta\Psi_A=&\Omega_{AB}\gamma^{\mu}\epsilon D_{\mu}X^B+\{\Omega_{AB}(\alpha_{2,2}X^CX_CX^B\\
&-\alpha_{2,1}X^BX_CX^C)-2\alpha_3\Omega_{BC}X^BX_AX^C\}\epsilon\\
\delta\Psi^A=&\Omega^{AB}\gamma^{\mu}\epsilon D_{\mu}X_B+\{\Omega^{AB}(-\alpha_{2,1}X_CX^CX_B\\
&+\alpha_{2,2}X_BX^CX_C)+2\alpha_3\Omega^{BC}X_BX^AX_C\}\epsilon\\
\delta A_{\mu}=&\frac{1}{k_1}[\Omega_{AB}\tilde{\epsilon}\gamma_{\mu}\Psi^AX^B+\Omega^{AB}X_B\tilde{\Psi}_A\gamma_{\mu}\epsilon]\\
\delta \hat{A}_{\mu}=&\frac{1}{k_2}[\Omega_{AB}X^B\tilde{\epsilon}\gamma_{\mu}\Psi^A+\Omega^{AB}\tilde{\Psi}_A\gamma_{\mu}\epsilon X_B]~,
\end{split}
\end{equation}
provided that the coefficients satisfy the relations:
\begin{equation}
\begin{split}
&a_1=-2i(\frac{1}{k_1}+\bar{\alpha}_1)~,~~a_2=2i(\frac{1}{k_2}+\alpha_1)~,\\
&a_3=-\bar{a}_3-i(\alpha_1-\bar{\alpha}_1)~,~~a_4=i(\alpha_1-\bar{\alpha}_1)~,\\
&\alpha_{2,1}=-\frac{1}{k_1}-2\bar{\alpha}_1~,~~\alpha_{2,2}=-\frac{1}{k_2}-2\alpha_1~,~~\alpha_3=i\bar{a}_3-\alpha_1~,\\
&\alpha_{4,1}=-3\alpha^2_{2,2}+4\alpha_{2,2}\alpha_3+m~,~~\alpha_{4,2}=-3\alpha^2_{2,1}+4\alpha_{2,2}\alpha_3+m~,\\
&\alpha_{4,3}=\alpha_{2,2}\alpha_3+\frac{m}{4}~,~~\alpha_{4,4}=-\alpha_{2,1}\alpha_{2,2}+2\alpha_{2,2}\alpha_3+\frac{m}{2}~,\\
&\bar{m}=4(\alpha_{2,2}-\alpha_{2,1})\alpha_3+m~,~~n=4(\alpha_3-\alpha_{2,2})\alpha_3-m~.
\end{split}
\end{equation}
In addition to the CS levels $k_1$, $k_2$, the theory has four independent parameters. One can choose them to be $\alpha_1$, $\bar{\alpha}_1$, $\bar{a}_3$ and $m$.

\subsection{Off-Shell}

In the previous section we studied the on-shell formulation of the theory. 
However to carry out 
the localization procedure one needs off-shell supersymmetry. 
For that purpose we introduce the 
auxiliary scalar fields $F$ and the gaugini $\lambda$, $\hat{\lambda}$
 in the scalar and gauge multiplets, respectively. The off-shell action reads:
\begin{equation}
\label{r19}
\cl=\cl_{CS}+\cl_{kin}+\cl_{\text{potential}}~, 
\end{equation}
where:
\begin{equation}
\begin{split}
\cl_{CS}=&\frac{k_1}{2\pi}\mathrm{tr}\left\{\varepsilon^{\mu\nu\rho}
(\frac{1}{2}A_{\mu}\partial_{\nu}A_{\rho}+\frac{i}{3}A_{\mu}A_{\nu}A_{\rho})+\frac{i}{2}\tilde{\lambda}\lambda \right\} \\
&-\frac{k_2}{2\pi}\mathrm{tr}\left\{\varepsilon^{\mu\nu\rho}(
\frac{1}{2}\hat{A}_{\mu}\partial_{\nu}\hat{A}_{\rho}
+\frac{i}{3}\hat{A}_{\mu}\hat{A}_{\nu}\hat{A}_{\rho})+\frac{i}{2}\tilde{\hat{\lambda}}\hat{\lambda}
  \right\}~,
\end{split}
\end{equation}
\begin{equation}
\cl_{kin}=\frac{1}{2\pi}\mathrm{tr}\left\{
-D^{\mu}X^AD_{\mu}X_A+i\tilde{\Psi}_A\gamma^{\mu}D_{\mu}\Psi^A-F^AF_A
\right\}~,
\end{equation}
\begin{equation}
\begin{split}
\cl_{\text{potential}}=&\frac{1}{2\pi}\mathrm{tr}\{i[(-\alpha_{2,1}X_BX^BX_A+\alpha_{2,2}X_AX^BX_B)-2\alpha_3\Omega_{AB}\Omega^{CD}X_CX^BX_D]F^A \\
&+iF_A[(-\alpha_{2,1}X^AX_BX^B+\alpha_{2,2}X^BX_BX^A)+2\alpha_3\Omega^{AB}\Omega_{CD}X^CX_BX^D]\} \\
&+\frac{1}{2\pi}\mathrm{tr}\{\Omega_{AB}\tilde{\lambda}\Psi^A
X^B-\Omega^{AB}X_B\tilde{\Psi}_A\lambda-\Omega_{AB}X^B\tilde{\Psi}^A\hat{\lambda}+\Omega^{AB}\tilde{\hat{\lambda}}\Psi_AX_B\} \\
&+\frac{1}{2\pi}\mathrm{tr}\{i\alpha_{2,1}\Omega^{AD}\Omega_{BC}\tilde{\Psi}_A\Psi^BX^CX_D-i\alpha_{2,2}\Omega_{AD}\Omega^{BC}\tilde{\Psi}^A\Psi_BX_CX^D \\
&-\frac{i}{2}\alpha_{2,2}\Omega^{AB}\Omega^{CD}\tilde{\Psi}_AX_B\Psi_CX_D+\frac{i}{2}\alpha_{2,1}\Omega_{AB}\Omega_{CD}\tilde{\Psi}^AX^B\Psi^CX^D \\
&+i\alpha_3\Omega^{AC}\Omega^{BD}\tilde{\Psi}_AX_B\Psi_CX_D-i\alpha_3\Omega_{AC}\Omega_{BD}\tilde{\Psi}^AX^B\Psi^CX^D \\
&-\frac{i}{2}\alpha_{2,1}\Omega^{AD}\Omega^{BC}\tilde{\Psi}_AX_B\Psi_CX_D+\frac{i}{2}\alpha_{2,2}\Omega_{AD}\Omega_{BC}\tilde{\Psi}^AX^B\Psi^CX^D\} \\
&+\frac{1}{2\pi}i\mathrm{tr}\{\alpha_{2,1}\tilde{\Psi}^A\Psi_AX_BX^B-\alpha_{2,2}\tilde{\Psi}_A\Psi^AX^BX_B\} \\
&+\frac{1}{2\pi}2i\mathrm{tr}\{\alpha_3\tilde{\Psi}_A\Psi^BX^AX_B-\alpha_3\tilde{\Psi}^B\Psi_AX_BX^A\}~.
\end{split}
\end{equation}
This can be rewritten compactly in superspace formalism, see e.g. (3.8) of \cite{Gaiotto:2009mv} which we reproduce here:
\eq{
\spl{
S=&\frac{k_1}{2\pi}S_{CS}({A})-\frac{k_2}{2\pi}S_{CS}(\hat{{A}})+\frac{1}{2\pi}\int d^2\theta\mathrm{tr}\{D_a\Phi^\dag_AD^a\Phi^A\\
&+(c_1\Phi^\dag_A\Phi^A\Phi^\dag_B\Phi^B+c_2\Phi^\dag_A\Phi^B\Phi^\dag_B\Phi^A+c_3\Omega^{AB}\Omega_{CD}\Phi^\dag_A\Phi^C\Phi^\dag_B\Phi^D)\}~,
}
}
where $\Phi_A$ is a superfield, and the connection with the component formulation discussed previously is provided by the relations:
\eq{
\label{r18}
c_1=-i\bar{\alpha}_1-\frac{i}{2k_1}~;~~~c_2=i\alpha_1+\frac{i}{2k_2}~;~~~c_3= i\alpha_1+\bar{a}_3~.
}
The action is invariant under the off-shell supersymmetry transformations:
\eq{
\spl{
\delta X_A&=i\Omega_{AB}\tilde{\epsilon}\Psi^B \\
\delta X^A&=i\Omega^{AB}\tilde{\epsilon}\Psi_B \\
\delta\Psi_A&=\Omega_{AB}\gamma^{\mu}\epsilon D_{\mu}X^B-i\Omega_{AB}F^B\epsilon \\
\delta\Psi^A&=\Omega^{AB}\gamma^{\mu}\epsilon D_{\mu}X_B-i\Omega^{AB}F_B\epsilon \\
\delta F_A&=-\Omega_{AB}\tilde{\epsilon}\gamma^\mu D_\mu\Psi^B-iX_A(\tilde{\epsilon}\hat{\lambda})+i(\tilde{\epsilon}\lambda)X_A \\
\delta F^A&=-\Omega^{AB}\tilde{\epsilon}\gamma^\mu D_\mu\Psi_B-iX^A(\tilde{\epsilon}\lambda)+i(\tilde{\epsilon}\hat{\lambda})X^A \\
\delta A_{\mu}&=-i\tilde{\epsilon}\gamma_\mu\lambda \\
\delta \hat{A}_{\mu}&=-i\tilde{\epsilon}\gamma_\mu\hat{\lambda} \\
\delta \lambda&=-\frac{1}{2}\gamma^{\mu\nu}\epsilon F_{\mu\nu} \\
\delta \hat{\lambda}&=-\frac{1}{2}\gamma^{\mu\nu}\epsilon \hat{F}_{\mu\nu}~.
}
}
We note that besides $k_1$, $k_2$ the off-shell theory has only three free parameters, as can be seen from (\ref{r18}). This is one fewer parameter than in the on-shell formulation. Specifically, after replacing the auxiliary field $F$ and gaugini $\lambda$, $\hat{\lambda}$ by the solutions of their  
respective equations of motion, the Lagrangian (\ref{r19}) goes back to (\ref{r20}), but with $\alpha_{4,3}=0$ in $\cl_{pot}$. In other words, for the on-shell theory obtained by starting from (\ref{r19}) and then eliminating the auxiliary fields, $m$ is not an independent parameter but is equal to $-4\alpha_{2,2}\alpha_3$, which in its turn can be expressed in terms of  $\alpha_1$, $\bar{\alpha}_1$ and $\bar{a}_{i}$. This can be  understood from the fact that the sextic potential $X_AX^BX_CX^AX_BX^C$ in $\cl_{pot}$ cannot be obtained from the off-shell Lagrangian by replacing $F$ by its solution.

In the following we will put  the theory on a curved manifold.
More specifically,  to go from flat to curved spacetime one needs to:
\begin{itemize}
\item covariantize all derivatives,
\item introduce additional terms $\frac{1}{3} \Omega_{AB}X^B\gamma^\mu\nabla_\mu \epsilon$ and $\frac{1}{3} \Omega^{AB}X_B\gamma^\mu\nabla_\mu \epsilon$ in the transformations of $\Psi_A$ and $\Psi^A$, respectively,
\item have $\epsilon$ satisfy the conformal Killing spinor equation:
\eq{\label{Killing}\nabla_\mu \epsilon = \gamma_\mu \eta~,} 
where $\eta$ is some arbitrary spinor,
\item add a scalar-curvature coupling term, $-\frac{1}{8}R X^A X_A$, to the Lagrangian.
\end{itemize}
Explicitly:
\begin{equation}
\begin{split}
\delta\Psi_A&\rightarrow\delta\Psi_A=\Omega_{AB}\gamma^{\mu}\epsilon D_{\mu}X^B+\frac{1}{3} \Omega_{AB}X^B\gamma^\mu\nabla_\mu \epsilon-i\Omega_{AB}F^B\epsilon~, \\
\delta\Psi^A&\rightarrow\delta\Psi^A=\Omega^{AB}\gamma^{\mu}\epsilon D_{\mu}X_B+\frac{1}{3} \Omega^{AB}X_B\gamma^\mu\nabla_\mu \epsilon-i\Omega^{AB}F_B\epsilon~,
\end{split}
\end{equation}
\begin{equation}
\cl_{kin}\rightarrow\cl_{kin}=\frac{1}{2\pi}\mathrm{tr}\left\{
-D^{\mu}X^AD_{\mu}X_A-\frac{1}{8}RX^AX_A+i\tilde{\Psi}_A\gamma^{\mu}D_{\mu}\Psi^A-F^AF_A
\right\}~.
\end{equation}
The resulting curved-space Lagrangian will be used in the next section.

\section{Localization}\label{sec:loc}

In order to apply the localization procedure, the theory must be invariant under the action of a fermionic symmetry $\delta$ which is nilpotent, $\delta^2=0$, or more generally squares to a symmetry of the theory. Deforming the action by a $\delta$-exact term,
\eq{
S\longrightarrow S+ t\delta V
~,}
leaves invariant the expectation values of $\delta$-closed operators. Hence we may take the limit $t\rightarrow \infty$, upon which the theory localizes to the set $\Sigma$ of critical points  of $\delta V$ \cite{wittentop}. 
In this limit the path integral can be performed by restricting $S$ to $\Sigma$ and computing a one-loop determinant describing the fluctuations normal to $\Sigma$. This procedure was first carried out in detail in \cite{Pestun:2007rz} for the case of SYM on the round $S^4$. 

In order for the path integral to be well-defined, we will consider the theory in Euclidean signature. All fields are then complexified, while the action becomes a holomorphic functional in the space of complexified fields. This procedure is known under the name of ``holomorphic complexification'' and ensures that supersymmetry is preserved, see e.g. \cite{Bergshoeff:2007cg}. Following \cite{Pestun:2007rz} our strategy will be to choose a path-integration contour in the space of fields, such that when restricted to that contour the deformation  $\delta V$ becomes a sum of positive semi-definite terms. The locus $\Sigma$ will then be determined by the condition that each term in the sum vanishes.

\subsection{Setup}

As explained above, in order to apply the localization procedure we need to pass from Lorentzian to Euclidean signature, where all fields become complex. Moreover $\epsilon^{\mu\nu\rho}$  in the CS piece of the Lagrangian becomes $i\epsilon^{\mu\nu\rho}$.

We then deform the action by adding a term $t\delta V$ such that $\delta^2V=0$. For theories with $\mathcal{N}\geqslant2$ supersymmetry, one can have $\delta^2=0$ on all fields of the theory. However, this is not possible for the  $\mathcal{N}=1$ superalgebra. Instead, as we will show later, for $\mathcal{N}=1$ we can require that $\delta$ squares to a transformation in the isometry group of the manifold, which in turn leads to $\delta^2V=0$ upon  volume integration. 

Furthermore we must restrict the supersymmetry parameter $\epsilon$ to satisfy the Killing spinor equation:\footnote{A detailed analysis of this Killing spinor equation in Lorentzian signature is given in section 3 of \cite{Andringa:2009yc}.}
\begin{equation}
\label{r8}
\nabla_\mu \epsilon = S \gamma_\mu \epsilon~,
\end{equation}
where $S$ is in general a complex function. 
The reason for restricting to this Killing spinor equation instead of the more general one (\ref{Killing}) is the following. Equation (\ref{Killing}) would in general imply that $\delta^2$ induces not only a translation, a rotation and a gauge transformation but also a dilatation, which would break the invariance of the deformation $\delta V$.

Under the assumption of smoothness, any solution to the Killing spinor equation which is not identically zero is nowhere-vanishing on the manifold. This follows from the fact that (\ref{r8}) is a first-order differential equation, hence if  the Killing spinor vanishes at any one point it must vanish everywhere.

Given a nowhere-vanishing Killing spinor $\epsilon$, any spinor $\Psi$ can be  decomposed as follows:
\begin{equation}
\Psi = \Psi_+ \epsilon + \Psi_-\epsilon^c~,
\end{equation}
where $\Psi_{\pm}$  are anticommuting scalars; our conventions are explained in appendix C. 
From now on we require the supersymmetry parameters to be commuting. The off-shell Lagrangian given in section 2 remains invariant under supersymmetry with these commuting parameters. With the above definitions the supersymmetric transformations can be rewritten as:
\begin{equation}
\begin{split}
\delta X_A=&ia\Omega_{AB}\Psi^B_-\\
\delta X^A=&ia\Omega^{AB}\Psi_{B-}\\
\delta \Psi_{A-}=&\frac{1}{a}\Omega_{AB}V^\mu D_\mu X^B\\
\delta \Psi_{A+}=&\frac{1}{a}\Omega_{AB}U^\mu D_\mu X^B+S\Omega_{AB}X^B-i\Omega_{AB}F^B\\
\delta \Psi^A_-=&\frac{1}{a}\Omega^{AB}V^\mu D_\mu X_B\\
\delta \Psi^A_+=&\frac{1}{a}\Omega^{AB}U^\mu D_\mu X_B+S\Omega^{AB}X_B-i\Omega^{AB}F_B\\
\delta F_A=&-\Omega_{AB}V^\mu D_\mu\Psi^B_++\Omega_{AB}U^\mu D_\mu\Psi^B_-\\
&+3S^\ast a\Omega_{AB}\Psi^B_--iaX_A\hat{\lambda}_-+ia\lambda_-X_A\\
\delta F^A=&-\Omega^{AB}V^\mu D_\mu\Psi_{B+}+\Omega^{AB}U^\mu D_\mu\Psi_{B-}\\
&+3S^\ast a\Omega^{AB}\Psi_{B-}-iaX^A\lambda_-+ia\hat{\lambda}_-X^A~,
\end{split}
\end{equation}
and for the gauge multiplets:
\begin{equation}
\begin{split}
\delta A_\mu=&-iV_\mu\lambda_++iU_\mu\lambda_-\\
\delta \hat{A}_\mu=&-iV_\mu\hat{\lambda}_++iU_\mu\hat{\lambda}_-\\
\delta \lambda_+=&-\frac{1}{2a}i\epsilon^{\mu\nu\rho}U_\rho F_{\mu\nu}\\
\delta \lambda_-=&-\frac{1}{2a}i\epsilon^{\mu\nu\rho}V_\rho F_{\mu\nu}\\
\delta \hat{\lambda}_+=&-\frac{1}{2a}i\epsilon^{\mu\nu\rho}U_\rho\hat{F}_{\mu\nu}\\
\delta \hat{\lambda}_-=&-\frac{1}{2a}i\epsilon^{\mu\nu\rho}V_\rho\hat{F}_{\mu\nu}~,
\end{split}
\end{equation}
where:
\begin{equation}
\begin{split}
&a\equiv \epsilon^\dag\epsilon=\tilde{\epsilon}\epsilon^c=-\tilde{\epsilon^c}\epsilon~,~~V^\mu\equiv\tilde{\epsilon}\gamma^\mu\epsilon~,\\
&U^\mu\equiv\epsilon^\dag\gamma^\mu\epsilon=-\tilde{\epsilon}\gamma^\mu\epsilon^c~,~~\nabla_\mu \epsilon^c=-S^\ast\gamma_\mu\epsilon^c~.
\end{split}
\end{equation}
Note that $\Psi^A_-$, $\Psi^A_+$, $\Psi_{A-}$, $\Psi_{A+}$, $\lambda_-$ and $\lambda_+$ are anticommuting; so is the supersymmetry transformation $\delta$. 
With the above setup, we find:
\begin{equation}
\begin{split}
\delta^2 X_A&=-iV^\mu D_\mu X_A\\
\delta^2 \Psi^A_-&=-iV^\mu D_\mu \Psi^A_-\\
\delta^2 \Psi^A_+&=-iV^\mu D_\mu \Psi^A_+ - 2ia(S-S^\ast)\Psi^A_-\\
\delta^2 F_A&=-V^\mu D_\mu F_A+V^\mu\partial_\mu SX_A~,\\
\end{split}
\end{equation}
and:
\begin{equation}
\begin{split}
\delta^2 A_\mu&=-iV^\nu F_{\nu\mu}\\
\delta^2 \hat{A}_\mu&=-iV^\nu \hat{F}_{\nu\mu}\\
\delta^2 \lambda_-&=-iV^\mu D_\mu \lambda_-\\
\delta^2 \lambda+&=-iV^\mu D_\mu \lambda_+ - 2ia(S-S^\ast)\lambda_-\\
\delta^2 \hat{\lambda}_-&=-iV^\mu D_\mu \hat{\lambda}_-\\
\delta^2 \hat{\lambda}+&=-iV^\mu D_\mu \hat{\lambda}_+ - 2ia(S-S^\ast)\hat{\lambda}_-~.
\end{split}
\end{equation}
Equivalently, written in terms of the original fields, two supersymmetry transformations give:
\begin{equation}
\label{r4}
\begin{split}
\delta^2X_A&=-iV^\mu D_\mu X_A\\
\delta^2\Psi^A&=-iV^\mu D_\mu \Psi^A-iSV^\mu\gamma_\mu\Psi^A\\
\delta^2F_A&=-iV^\mu D_\mu F_A+V^\mu\partial_\mu SX_A~,
\end{split}
\end{equation}
and:
\begin{equation}
\label{r9}
\begin{split}
\delta^2A_\mu&=-iV^\nu F_{\nu\mu}\\
\delta^2\hat{A}_\mu&=-iV^\nu \hat{F}_{\nu\mu}\\
\delta^2\lambda&=-iV^\mu D_\mu \lambda-iSV^\mu\gamma_\mu\lambda\\
\delta^2\hat{\lambda}&=-iV^\mu D_\mu \hat{\lambda}-iSV^\mu\gamma_\mu \hat{\lambda}~.\\
\end{split}
\end{equation}
As explained in Appendix C, $V^\mu$ can be identified as part of the orthonormal frame that trivializes the tangent bundle of the manifold. Therefore, apart from additional terms which can be interpreted as gauge transformations or rotations, $\delta^2$ acting on each field gives a translation along $V^\mu$.

In the next section we will ultimately set $a=1$ and $S=0$, upon which 
the above equations simplify further.

\subsection{Deformations}

\subsubsection{Matter Sector}

To localize the matter sector, we first consider the deformation,  
\eq{\label{r}\delta V=\int\sqrt{g}d^3x\delta[(\delta\Psi_A)^\dag\Psi_A]~,}
where we have {\it defined}:
\begin{equation}\label{r5}
(\delta\Psi_A)^\dag\equiv\Omega^{AB}\epsilon^\dag\gamma^\mu D_\mu X_B+S^\ast\Omega^{AB}X_B\epsilon^\dag+i\Omega^{AB}F_B\epsilon^\dag~.
\end{equation}
Note that at generic points in field space $(\delta\Psi_A)^\dag$ is {\it not} the adjoint of $\delta\Psi_A$, and $\delta V$ as defined in (\ref{r}) is a holomorphic functional in the space of complexified fields.

As explained in section \ref{sec:loc}, we will choose a path-integration contour $\mathcal{C}$ in the space of fields such that when restricted to $\mathcal{C}$ the deformation  $\delta V$ becomes a sum of positive semi-definite terms. This requirement selects $\mathcal{C}$ as the subspace where the 
fields satisfy the reality condition:
\eq{\label{r12}
\mathrm{Contour}~\mathcal{C}:~~~
\begin{array}{rl}
X^{A\dag}=X_A~,&F^{A\dag}=F_A~,\\
A^\dag_\mu=A_\mu~,&{\hat{A}_\mu}^\dag=\hat{A}_\mu~.
\end{array}
}
Moreover the integrand in (\ref{r}) is given by:
\begin{equation}
\label{r3}
\delta[(\delta\Psi_A)^\dag\Psi_A]=\delta(\delta\Psi_A)^\dag\Psi_A+(\delta\Psi_A)^\dag\delta\Psi_A~.
\end{equation}
Recall that the supersymmetry transformation $\delta$ is anticommuting; the  relative sign on the right-hand side is positive since $(\delta\Psi_A)^\dag$ is bosonic.

Let us now verify that the deformation is $\delta$-closed. From ($\ref{r3}$) 
we obtain:
\begin{equation}
\begin{split}
\delta^2[(\delta\Psi_A)^\dag\Psi_A]=&\delta^2(\delta\Psi_A)^\dag\Psi_A-\delta(\delta\Psi_A)^\dag\delta\Psi_A+\delta(\delta\Psi_A)^\dag\delta\Psi_A+(\delta\Psi_A)^\dag\delta^2\Psi_A \\
=&\delta^2(\delta\Psi_A)^\dag\Psi_A+(\delta\Psi_A)^\dag\delta^2\Psi_A~.
\end{split}
\end{equation}
The second term in the second line can be read off from (\ref{r4}). One can obtain the first term in the second line from (\ref{r4}) and (\ref{r5}):
\begin{equation}
\begin{split}
\delta^2(\delta\Psi_A)^\dag\Psi_A=&-iV^\mu D_\mu[(\delta\Psi_A)^\dag]\Psi_A+iS^\ast V^\mu(\delta\Psi_A)^\dag\gamma_\mu\Psi_A\\
&+2iS^\ast\Omega^{AB}V_\mu D_\nu X_B\epsilon^\dag\gamma^{\mu\nu}\Psi_A-2iS\Omega^{AB}V_\mu D_\nu X_B\epsilon^\dag\gamma^{\mu\nu}\Psi_A~,
\end{split}
\end{equation}
where we used $\nabla_\mu\epsilon^c=-S^\ast\gamma_\mu\epsilon^c$ and chose $S$ to be a constant. 
Finally,
\begin{equation}
\label{r6}
\begin{split}
\delta^2[(\delta\Psi_A)^\dag\Psi_A]=&-iV^\mu\partial_\mu [(\delta\Psi_A)^\dag\Psi_A] \\
&+iS^\ast V^\mu(\delta\Psi_A)^\dag\gamma_\mu\Psi_A-iSV^\mu(\delta\Psi_A)^\dag\gamma_\mu\Psi_A\\
&+2iS^\ast\Omega^{AB}V_\mu D_\nu X_B\epsilon^\dag\gamma^{\mu\nu}\Psi_A-2iS\Omega^{AB}V_\mu D_\nu X_B\epsilon^\dag\gamma^{\mu\nu}\Psi_A~.
\end{split}
\end{equation}
This vanishes under the volume integration if and only if $S$ is real constant. 
On the other hand the integrability condition of the Killing spinor (\ref{r8}) relates the constant $S$ to the curvature scalar of the manifold:
\eq{
R=-24S^2~.
}
If $S$ is nonvanishing, this would allow hyperbolic space as a solution. In the following we will discard this possibility and instead demand that the manifold should be compact, in order to  ensure that the partition function is well-defined.

On $T^3$, the curvature scalar vanishes and so does $S$. This implies that the Killing spinor is constant and  nowhere-vanishing. Moreover, in (\ref{r4}) and (\ref{r9}), with vanishing $S$ terms, $\delta^2$ gives a translation and a gauge transformation on all fields. $\delta$-exactness and $\delta$-closedness of the deformation are thus guaranteed.

We will henceforth restrict the manifold to be $T^3$. 
We normalize the constant Killing spinor such that $\tilde{\epsilon}\epsilon^c=1$. The bosonic part of the deformation (\ref{r3}) is:
\begin{equation}
\begin{split}
(\delta\Psi_A)^\dag\delta\Psi_A=&D_\mu X_AD^\mu X^A+i\epsilon^{\mu\nu\rho}U_\rho D_\mu X_AD_\nu X^A+F_AF^A\\
&+iU^\mu D_\mu X^AF_A-iU^\mu D_\mu X_AF^A~,
\end{split}
\end{equation}
where $U_\mu$ is a real unit vector, which we may choose to be along the third direction of $T^3$ without loss of generality. When restricted to the contour $\mathcal{C}$, cf.~(\ref{r12}), the bosonic part of the deformation is positive semi-definite, and the saddle points where it vanishes are given by:
\begin{equation}
\label{pl}
D_1X_A+iD_2X_A=0~,~~D_3X-iF=0~.
\end{equation}
Hence with this deformation alone the theory does not reduce to an ordinary  integral with discrete saddle points: one can always choose some nontrivial functions for $X_A$ and $F$ so that (\ref{pl}) is satisfied. We 
therefore add another term $\delta[(\delta\Psi^A)^\dag\Psi^A]$ to the original deformation:
\begin{equation}
\begin{split}
(\delta\Psi_A)^\dag\delta\Psi_A+(\delta\Psi^A)^\dag\delta\Psi^A
=&D_\mu X_AD^\mu X^A+i\epsilon^{\mu\nu\rho}U_\rho D_\mu X_AD_\nu X^A+F_AF^A\\
&+iU^\mu D_\mu X^AF_A-iU^\mu D_\mu X_AF^A\\
&+D_\mu X_AD^\mu X^A-i\epsilon^{\mu\nu\rho}U_\rho D_\mu X_AD_\nu X^A+F_AF^A\\
&-iU^\mu D_\mu X^AF_A+iU^\mu D_\mu X_AF^A\\
=&2\{D_\mu X_AD^\mu X^A+F_AF^A\}~.
\end{split}
\end{equation}
When restricted to the contour $\mathcal{C}$, 
the two terms in the last line are both positive semi-definite, and the critical points are given by:
\begin{equation}
D_\mu X_A=F_A=0~.
\end{equation}

\subsubsection{Gauge Sector}

A $\delta$-closed deformation for the gauge sector is:
\begin{equation}
\label{r11}
\int d^3x\left\{\delta[(\delta{\lambda})^\dag\lambda]
+\delta[(\delta\hat{\lambda})^\dag\hat{\lambda}]
\right\}~,
\end{equation}
where we have {\it defined}:
\eq{(\delta{\lambda})^\dag\equiv -\epsilon^\dag\gamma_{\mu\nu}{F}^{\mu\nu}~;~~~
(\delta\hat{\lambda})^\dag\equiv-\epsilon^\dag\gamma_{\mu\nu}\hat{F}^{\mu\nu}~,}
so that the deformation (\ref{r11})  is a holomorphic functional of the complexified fields. 
Note in particular that $(\delta{\lambda})^\dag$ is {\it not} the adjoint of $\delta{\lambda}$ at generic points in field space, but only when restricted to the contour $\mathcal{C}$, cf.~(\ref{r12}). 

The bosonic part of the deformation (\ref{r11}) is given by:
\begin{equation}
(\delta{\lambda})^\dag \delta\lambda
+(\delta\hat{\lambda})^\dag \delta\hat{\lambda}
=\frac{1}{2}F^{\mu\nu}F_{\mu\nu}+\frac{1}{2}\hat{F}^{\mu\nu}\hat{F}_{\mu\nu}~.
\end{equation}
When restricted to the contour $\mathcal{C}$ this becomes a sum of positive semi-definite terms, with critical points given by:
\eq{{F}_{\mu\nu}=\hat{F}_{\mu\nu}=0~.}

\subsection{Gauge Fixing}

We now introduce  the usual ghost and anti-ghost action to fix the infinite degrees of freedom of the gauge fields. The ghost term is not invariant under supersymmetry, so one cannot immediately proceed to do localization. To deal with this, we follow \cite{Pestun:2007rz,Kapustin:2009kz}, and introduce a new fermionic symmetry $\Delta$:
\begin{equation}
\Delta\equiv\delta_Q+\delta_B~,
\end{equation}
where $\delta_Q$ stands for supersymmetry and $\delta_B$ for BRST transformation.

Under a BRST transformation, we have:
\begin{equation}
\delta_BA_\mu=\partial_\mu C+i[A_\mu,C]~,~~\delta_B\lambda=-i\{\lambda,C\}~.
\end{equation}
and similarly  for $\hat{A}$, $\hat{\lambda}$. 
Here $C$ is the usual anti-commuting ghost field. It transforms under supersymmetry and BRST as:
\begin{equation}
\delta_QC=0~,~~\Delta C=\delta_BC=a_0-\frac{i}{2}\{C,C\}~,~~\Delta a_0=0~,
\end{equation}
where $a_0$ is a constant ghost-for-ghost field that takes care of the zero mode of $C$. With this combined transformation, one can verify that:
\begin{equation}
\begin{split}
\Delta^2 A_\mu=&-iV^\nu F_{\nu\mu}+i[A_\mu,a_0]~,\\
\Delta^2 \lambda=&-iV^\mu D_\mu\lambda+i[\lambda,a_0]~,\\
\Delta^2 C=&i[C,a_0]~.
\end{split}
\end{equation}
The rest of the ghost complex transforms under $\Delta$ as:
\begin{equation}
\begin{split}
&\Delta \bar{C}=b~,~~\Delta b=-iV\cdot D\bar{C}+i[\bar{C},a_0]~,\\
&\Delta \bar{a}_0=\bar{C}_0~,~~\Delta\bar{C}_0=i[\bar{a}_0,a_0]~,\\
&\Delta b_0=C_0~,~~\Delta C_0=[V\cdot A,b_0]+[A_\mu,\partial^\mu(V\cdot A)]-i\DAlambert(V\cdot A)+i[b_0,a_0]~,
\end{split}
\end{equation}
where $\bar{C}$ is the anti-ghost, and $b$ is the Lagrangian multiplier; $\bar{a}_0$,$b_0$,$C_0$ and $\bar{C}_0$ are constant fields needed to fix the zero modes of the ghosts and $b$.

The gauge-fixing action is:
\begin{equation}
\label{r7}
\begin{split}
&i\int d^3x \mathrm{tr}\{\Delta[\bar{C}(\partial^\mu A_\mu+b_0)-C\bar{a}_0]\}\\
=&i\int d^3x \mathrm{tr}\{b(\partial^\mu A_\mu+b_0)-\bar{C}(\partial^\mu D_\mu C+\partial^\mu\delta_QA_\mu+C_0)\\
&-(a_0-\frac{i}{2}\{C,C\})\bar{a}_0+C\bar{C}_0\}~.
\end{split}
\end{equation}
Note that the ghost, the anti-ghost and the  transformation $\Delta$ are all anti-commuting. In Appendix D we show that the integration over all fields in the ghost complex gives the Lorentz gauge. 
Now this action is invariant under $\Delta$ transformation:
\begin{equation}
\begin{split}
&\Delta^2[\bar{C}(\partial^\mu A_\mu+b_0)-C\bar{a}_0]\\
=&\Delta^2(\bar{C})(\partial^\mu A_\mu+b_0)+\bar{C}(\partial^\mu\Delta^2(A_\mu)+\Delta^2(b_0))\\
&-\Delta^2(C)\bar{a}_0-C\Delta^2(\bar{a}_0)\\
=&\Delta^2(\bar{C})(\partial^\mu A_\mu+b_0)+\bar{C}(\partial^\mu\Delta^2(A_\mu)+\Delta^2(b_0))\\
&-i[C,a_0]\bar{a}_0-iC[\bar{a}_0,a_0]~.
\end{split}
\end{equation}
The last two terms cancel under the trace. The first two can also be shown to cancel:
\begin{equation}
\begin{split}
&\int d^3x \mathrm{tr}\{\Delta^2(\bar{C})(\partial^\mu A_\mu+b_0)+\bar{C}(\partial^\mu\Delta^2(A_\mu)+\Delta^2(b_0))\}\\
=&\int d^3x \mathrm{tr}\{(-iV\cdot D\bar{C}+i[\bar{C},a_0])(\partial^\mu A_\mu+b_0)\\
&+\bar{C}\partial^\mu(-iV^\nu F_{\nu\mu}+i[A_\mu,a_0])\\
&+\bar{C}([V\cdot A,b_0]+[A_\mu,\partial^\mu(V\cdot A)]-i\DAlambert(V\cdot A)+i[b_0,a_0])\}\\
=&\int d^3x \mathrm{tr}\{i[\bar{C},a_0](\partial\cdot A+b_0)+i\bar{C}([\partial\cdot A+b_0,a_0])\\
&-iV\cdot\partial[\bar{C}(\partial\cdot A+b_0)]+[V\cdot A,\bar{C}](\partial\cdot A+b_0)\\
&+\bar{C}[V\cdot A,\partial\cdot A+b_0]+i\bar{C}\DAlambert(V\cdot A)-i\bar{C}\DAlambert(V\cdot A)\\
&\bar{C}[\partial^\mu(V\cdot A),A_\mu]+\bar{C}[A_\mu,\partial^\mu(V\cdot A)]\}\\
=&0~.
\end{split}
\end{equation}

\subsection{Saddle Points}\label{sec:sad}

For the gauge sector we replace $\delta$ by $\Delta$ in (\ref{r11}) and modify the deformation as follows:
\eq{
\spl{
\Delta V_\text{gauge}=&\int dx^3\Delta\mathrm{tr}\{\frac{1}{2}\epsilon^\dag\gamma_{\mu\nu}F_{\mu\nu}\lambda\}\\
=&\int dx^3\mathrm{tr}\{\frac{1}{2}F_{\mu\nu}F^{\mu\nu}-i\tilde{\lambda}\slashed{D}\lambda\}~,
}
}
and similarly for the hatted fields. 
This deformation is $\Delta$-exact and $\Delta$-closed. 
For the matter sector, $\Delta$ is defined to be the same as $\delta$, and the deformation is:
\eq{
\spl{
\Delta V_\text{matter}=&\int dx^3\mathrm{tr}\{\Delta[(\Delta \Psi^A)^\dag\Psi^A+(\Delta\Psi_A)^\dag\Psi_A]\}\\
=&2\int dx^3\mathrm{tr}\{D_\mu X_AD^\mu X^A+F_AF^A-i\tilde{\Psi}^A\slashed{D}\Psi_A\\
&+\Omega^{AB}\tilde{\lambda}X_B\Psi_A+\Omega_{AB}\tilde{\hat{\lambda}}X^B\Psi^A
-\Omega^{AB}X_B\tilde{\hat{\lambda}}\Psi_A-\Omega_{AB}X^B\tilde{\lambda}\Psi^A\}~.
}
}
The gauge sector localizes to:
\eq{\label{f}
F_{\mu\nu}=0~;~~~\lambda=0~,
}
where we have restricted to the contour $\mathcal{C}$, cf.~(\ref{r12}). 
In particular the saddle points of the gauge field correspond  to flat gauge connections over the Euclidean three-torus. For a simply-connected gauge group 
$\pi_1(G)=0$, such as $G=SU(N)\times SU(N)$, this implies that:
\eq{\label{s1}A_{\mu}=c^i_{\mu}H_i~,}
where $c^i$'s are constants and $\{H_i\}$, $i=1,\cdots,\mathrm{rank}(G)$, is the Cartan subalgebra of $G$. 
This can be seen as follows (see e.g. \cite{Witten:1982df,Keurentjes:1998uu}): Since $A_{\mu}$ is a flat connection there exists 
a group element $U\in G$ such that $A_{\mu}=-i\partial_{\mu}U U^{-1}$, at 
least locally. I.e. $U$ need not be globally defined but is allowed to undergo $G$-valued jumps as we wind around each of the three circles of the 
torus. More explicitly, suppose we have a square torus of radius $L$ parameterized by $\{x^{\mu}\in[0,L]\}$. The group element $U(x^1,x^2,x^3)$ 
obeys nontrivial, in general, boundary conditions which may be parameterized 
as follows,
\eq{\spl{\label{h1}U(x^1+L,x^2,x^3)&=U(x^1,x^2,x^3)\Omega_1~;\\
U(x^1,x^2+L,x^3)&=U(x^1,x^2,x^3)\Omega_2~;\\
U(x^1,x^2,x^3+L)&=U(x^1,x^2,x^3)\Omega_3~,}}
for some constant $\Omega_{\mu}\in G$. In addition, for consistency, $\Omega_{\mu}$ must mutally commute. Indeed going once around the circle parameterized by $x^{\mu}$ and then once around the circle parameterized by $x^{\nu}$ must produce the same jump in $U$ as when going first around the  $x^{\nu}$ direction and then along $x^{\mu}$. This implies, taking (\ref{h1}) into account, 
\eq{[\Omega_{\mu},\Omega_{\nu}]=0~.}
For a unitary group $G$, as is the case in the present paper, this implies that 
$\Omega_{\mu}$ can be put in the form:
\eq{
\Omega_{\mu}=\exp(iLc^j_{\mu}H_j)
~,}
up to similarity transformation. Recalling the relation between $A_{\mu}$ and 
$U$ we are thus led to the result cited in (\ref{s1}), provided we can show that for any set of mutally commuting $\Omega_{\mu}$'s we can always construct a group element $U\sim\exp(ix^{\mu}c^j_{\mu}H_j)$ obeying (\ref{h1}).

The proof of the last step proceeds by showing that there is no obstruction 
in constructing an element $U(x^{1}x^{2},x^{3})$ on the edges of a cube of side $L$ 
such that (\ref{h1}) is satisfied. Then $U$ can be continued on the faces of the cube provided $\pi_1(G)=0$, and finally in the interior provided 
 $\pi_2(G)=0$, which holds true for $G=SU(N)\times SU(N)$.

An important observation is that the constants $c^i_{\mu}$ should be understood as periodic variables with periodic identification,
\eq{\label{s2}c^i_{\mu}\sim c^i_{\mu}+\frac{2\pi}{L}~.}
This can be seen by performing a gauge transformation generated by 
$U=\exp(\frac{2\pi i}{L}x^{\mu}H_i)$, which shifts $A_{\mu}$ in accordance with  (\ref{s2}). On the other hand the element $U$ thus defined is periodic\footnote{We are adopting the normalization $\exp({2\pi i}H_i)=1$.},  i.e. as we wind around the $x^{\mu}$ direction of the torus it forms a closed 
loop in group space. But since the group is simply connected $U$ may be continuously deformed to the identity, and the gauge transformation generated by $U$ should act trivially on all fields of the theory. We thus arrive at the identification (\ref{s2}). 
 
It follows from the above that the $c^i_{\mu}$'s  can be constrained to take values in $[0,\frac{2\pi}{L}]$. 
In particular taking the infinite-volume limit of the 
torus, $L\rightarrow\infty$, we conclude that the only solution to (\ref{f}) 
is the trivial flat connection $A_{\mu}=0$. Of course on $\mathbb{R}^3$ there 
is no obstruction to gauging away any flat connection of the form (\ref{s1}). 
The point is that we can  
formally reproduce this result by 
considering $\mathbb{R}^3$ as the infinite-volume limit of $T^3$.

The case of $G=U(N)\times U(N)$ presents one crucial difference: 
 $\pi_1(U(N))\cong\mathbb{Z}$ and thus $G$ is not simply connected. 
By considering the decomposition of the algebra-valued connection along 
the $G$-generators it is not very difficult to see that we may still put the most general flat connection in the form (\ref{s1}),
\eq{\label{s}A_{\mu}=c^i_{\mu}H_i+d_{\mu}J+e_{\mu}K~,}
where the first term on the right-hand side is as in the case of $SU(N)\times SU(N)$;  $d_{\mu}$, $e_{\mu}$ are constants; $J$, $K$ are the two additional $u(1)$ Cartan generators coming from the decomposition:
\eq{u(N)\oplus u(N)\cong su(N)\oplus su(N)\oplus u(1)\oplus u(1)~.} 
Now the previous argument which allowed us to conclude that $c^i_{\mu}$ are 
periodic does not go through for the variables $d_{\mu}$, $e_{\mu}$. 
The reason is that the gauge transformations generated by $U=\exp(\frac{2\pi i}{L}x^{\mu}J)$ and $U=\exp(\frac{2\pi i}{L}x^{\mu}K)$ form closed loops in 
the group space which are not contractible to the identity. Hence the  gauge transformations generated by $U$ need not act trivially on all fields of the theory. 

In particular our argument that in the infinite-volume limit the only 
flat connection is the trivial one, does not go through in this case without  additional assumptions. If we wish to recover $A=0$ as the unique (up to 
gauge transformations) solution 
to (\ref{s}) in the infinite-volume limit, we must impose by hand that 
$U=\exp(\frac{2\pi i}{L}x^{\mu}J)$ and $U=\exp(\frac{2\pi i}{L}x^{\mu}K)$ act trivially on all fields of the theory.

Finally, the matter sector localizes to the following field configurations:
\eq{
F_A=0~;~~~\Psi_A=\Psi^A=0~;~~~X_A=\text{const}~,
}
where we have restricted to the contour $\mathcal{C}$, cf.~(\ref{r12}).

\subsection{One-loop Determinant}

We will now compute the one-loop determinant from the quadratic fluctuations around the following saddle points,
\eq{
\label{r22}
\spl{
&A_\mu=0~;~~~\lambda=0~;\\
&F_A=0~;~~~\Psi_A=\Psi^A=0~;~~~X_A=\text{const}~,
}
}
and similarly for $\hat{A}$, $\hat{\lambda}$.  
I.e. we will ignore the contributions from non-vanishing flat gauge connections, as discussed in the previous section.

The full path integral is of the form:
\eq{\label{pi}
\int d\varphi \exp\{iS+iS_\text{g.f.}-t(\Delta V_\text{gauge}+\frac{1}{2}\Delta V_\text{matter})\}~,
}
where $iS_\text{g.f.}$ is the gauge-fixing action (\ref{r7}), and $\int d\varphi$ stands for integrations over all fields and ghosts; $\Delta V_\text{gauge}$ contains deformations for both hatted and unhatted gauge multiplets.

Next we expand the fields around the saddle points:
\eq{
X_A\rightarrow X^0_A+\frac{1}{\sqrt{t}}X^\prime_A~,~~~\phi\rightarrow 0+\frac{1}{\sqrt{t}}\phi~.
}
Here $X^0_A$ is a constant field and $X^\prime_A$ represents the nonzero mode of $X_A$; $\phi$ stands for all fields other than $X_A$. The path integral (\ref{pi}) is $t$-independent 
thanks to localization. On the other hand, 
taking $t\rightarrow\infty$ allows us to keep only the quadratic terms in the deformation:
\eq{
\label{r13}
\spl{
&t(\Delta V_\text{gauge}+\frac{1}{2}\Delta V_\text{matter})\\
=&\int dx^3\mathrm{tr}\{\frac{1}{2}F^A_{\mu\nu}F^{A\mu\nu}-i\tilde{\lambda}\slashed{\partial}\lambda\}+\int dx^3\mathrm{tr}\{\frac{1}{2}\hat{F}^A_{\mu\nu}\hat{F}^{A\mu\nu}-i\tilde{\hat{\lambda}}\slashed{\partial}\hat{\lambda}\}\\
&+\int dx^3\mathrm{tr}\{\partial_\mu X^\prime_A\partial^\mu X^{\prime A}+X^{0A}A_\mu A^\mu X^0_A+X^{0}_A\hat{A}_\mu\hat{A}^\mu X^{0A}-2X^0_A\hat{A}_\mu X^{0A}A^\mu\\
&+F_AF^A-i\tilde{\Psi}^A\slashed{\partial}\Psi_A+\Omega^{AB}\tilde{\lambda}X^0_B\Psi_A+\Omega_{AB}\tilde{\hat{\lambda}}X^{0B}\Psi^A-\Omega^{AB}X^0_B\tilde{\hat{\lambda}}\Psi_A-\Omega_{AB}X^{0B}\tilde{\lambda}\Psi^A\}~,
}
}
where $F^A_{\mu\nu}\equiv\partial_{\mu}A_{\nu}-\partial_{\nu}A_{\mu}$ is the linearized field strength; some terms have been eliminated using Lorentz gauge.

\subsubsection{Determinant from Bosons}

We start with the calculation of the one-loop determinant of the bosonic part. 
Under Lorentz gauge, we have:
\eq{
\spl{
&\int d^3x\mathrm{tr}\{\frac{1}{2}F^A_{\mu\nu}F^{A\mu\nu}\}+\int dx^3\mathrm{tr}\{\frac{1}{2}\hat{F}_{A\mu\nu}\hat{F}^{A\mu\nu}\}+\int d^3x\mathrm{tr}\{\partial_\mu X^\prime_A\partial^\mu X^{\prime A}\\
&+X^{A0}A_\mu A^\mu X^0_A+X^{0}_A\hat{A}_\mu\hat{A}^\mu X^{A0}-2X^0_A\hat{A}_\mu X^{A0}A^\mu+F_AF^A\}\\
=&\int d^3x\mathrm{tr}\{-A_\mu\DAlambert A^\mu\}+\int dx^3\mathrm{tr}\{-\hat{A}_\mu\DAlambert\hat{A}^\mu\}+\int d^3x\mathrm{tr}\{-X^\prime_A\DAlambert X^{\prime A}\\
&+X^{A0}A_\mu A^\mu X^0_A+X^{0}_A\hat{A}_\mu\hat{A}^\mu X^{A0}-2X^0_A\hat{A}_\mu X^{A0}A^\mu+F_AF^A\}~.
}
}
On $T^3$ with periodic conditions, any field $\varphi$ can be expanded in terms of Fourier modes:
\eq{
\varphi=\sum_{\vec{n}}\varphi_{\vec{n}}\exp\{i2\pi\vec{n}\cdot\vec{x}\}~,
}
where $\vec{n}=(n_x,n_y,n_z)$ and each $n_\mu$ runs over all integers. 
In addition, for the gauge field the Lorentz gauge implies that for each $\vec{n}$, 
\eq{
n_xA_{x,\vec{n}}+n_yA_{y,\vec{n}}+n_zA_{z,\vec{n}}=0~.
}
Let us first assume $n_z\neq 0$. (We will come back to the case $n_z=0$ in the following). Then the previous equation can be used to eliminate $A_{z,\vec{n}}$ via:
\eq{
A_{z,\vec{n}}=-\frac{n_x}{n_z}A_{x,\vec{n}}-\frac{n_y}{n_z}A_{y,\vec{n}}~.
} 
The gauge fields are in the adjoint representation, $A_\mu=A^a_\mu t_a$, 
where the generators $t_a$ are normalized so that $\mathrm{tr}\{t_at_b\}=\delta_{ab}$. The gauge kinetic action becomes:
\eq{
\spl{
&\int d^3x\mathrm{tr}\{-A_\mu\DAlambert A^\mu\}\\
=&\int d^3x\sum_a\sum_{\vec{n},n_z\neq0}4\pi^2\vec{n}^2\{(\frac{n^2_x+n^2_z}{n^2_z})A^a_{x,-\vec{n}}A^a_{x,\vec{n}}+(\frac{n^2_y+n^2_z}{n^2_z})A^a_{y,-\vec{n}}A^a_{y,\vec{n}}\\
&+\frac{n_xn_y}{n^2_z}A^a_{x,-\vec{n}}A^a_{y,\vec{n}}+\frac{n_xn_y}{n^2_z}A^a_{y,-\vec{n}}A^a_{x,\vec{n}}\}~.
}
}
By symmetrizing $\vec{n}$ and $-\vec{n}$, for each pair of $(\vec{n},-\vec{n})$ and each $a$, this can be written in matrix notation as follows:
\eq{
\begin{array}{ccc}
&\begin{array}{cccc}
\ A^a_{x,\vec{n}}&\ A^a_{x,-\vec{n}}&\ A^a_{y,\vec{n}}&\ A^a_{y,-\vec{n}}
\end{array}&\\
\begin{array}{c}
A^a_{x,\vec{n}}\\
\ \ A^a_{x,-\vec{n}}\\
A^a_{y,\vec{n}}\\
\ \ A^a_{y,-\vec{n}}
\end{array}&
\begin{pmatrix}
0&\frac{n^2_x+n^2_z}{n^2_z}&0&\frac{n_xn_y}{n^2_z}\\
\frac{n^2_x+n^2_z}{n^2_z}&0&\frac{n_xn_y}{n^2_z}&0\\
0&\frac{n_xn_y}{n^2_z}&0&\frac{n^2_y+n^2_z}{n^2_z}\\
\frac{n_xn_y}{n^2_z}&0&\frac{n^2_yn^2_z}{n^2_z}&0
\end{pmatrix}&\times \ 4\pi^2(\vec{n}\cdot\vec{n})~.
\end{array}
}
Similarly, for each $(\vec{n},-\vec{n})$ and $a,b$, the potentials involving the gauge fields are:
\eq{
\spl{
X^{A0}A^a\cdot A^bt_at_bX^0_A:&~~~\Gamma\times X^{A0}t_at_bX^0_A~,\\
X^0_A\hat{A}^a\cdot\hat{A}^b\hat{t}_a\hat{t}_bX^{A0}:&~~~\Gamma\times X^0_A\hat{t}_a\hat{t}_bX^{A0}~,\\
-2X^0_A\hat{A}^a_\mu\hat{t}_aX^{A0}A^{b\mu}t_b:&~~~\Gamma\times -2X^0_A\hat{t}_aX^{A0}t_b~,
}
}
where:
\eq{
\Gamma\equiv\begin{pmatrix}
0&\frac{n^2_x+n^2_z}{n^2_z}&0&\frac{n_xn_y}{n^2_z}\\
\frac{n^2_x+n^2_z}{n^2_z}&0&\frac{n_xn_y}{n^2_z}&0\\
0&\frac{n_xn_y}{n^2_z}&0&\frac{n^2_y+n^2_z}{n^2_z}\\
\frac{n_xn_y}{n^2_z}&0&\frac{n^2_yn^2_z}{n^2_z}&0
\end{pmatrix}~.
}
The matter fields are in the bifundamental representation of the gauge group $U(N)\times U(N)$. Moreover  $X=\sum_{(\rho,\hat{\rho})}X^{(\rho,\hat{\rho})}\Ket{\rho}\otimes\Ket{\hat{\rho}}$, where $\Ket{\rho}$, $\Ket{\hat{\rho}}$ are representatives of the weights in each weight space; we choose the normalization so that $\braket{\rho|\rho^\prime}=\delta_{\rho,\rho^\prime}$ and $\braket{\hat{\rho}|\hat{\rho}^\prime}=\delta_{\hat{\rho},\hat{\rho}^\prime}$, in some gauge-invariant contraction of the relevant color indices. We then have:
\eq{
\spl{
X^{A0}t_at_bX^0_A=&\sum_{(\rho,\hat{\rho})}\sum_{(\rho^\prime,\hat{\rho}^\prime)}X^{A0(\rho,\hat{\rho})}\bra{\hat{\rho}}\otimes\bra{\rho}t_at_b\ket{\rho^\prime}\ket{\hat{\rho}^\prime}X^{0(\rho^\prime,\hat{\rho}^\prime)}_A\\
=&\sum_{\rho,\rho^\prime,\hat{\rho},\rho^{\prime\prime}}X^{A0(\rho,\hat{\rho})}\bra{\rho}t_a\ket{\rho^{\prime\prime}}\bra{\rho^{\prime\prime}}t_b\ket{\rho^\prime}X^{0(\rho^\prime,\hat{\rho})}_A\\
=&\sum_{\rho,\rho^\prime,\hat{\rho},\rho^{\prime\prime}}X^{A0(\rho,\hat{\rho})}\sigma^{(\rho,\rho^{\prime\prime})}_a\sigma^{(\rho^{\prime\prime},\rho^\prime)}_bX^{0(\rho^\prime,\hat{\rho})}_A~,\\
X^{0}_A\hat{t}_a\hat{t}_bX^{A0}=&\sum_{\hat{\rho},\hat{\rho}^\prime,\rho,\hat{\rho}^{\prime\prime}}X^{0(\rho,\hat{\rho})}_A\hat{\sigma}^{(\hat{\rho},\hat{\rho}^{\prime\prime})}_a\hat{\sigma}^{(\hat{\rho}^{\prime\prime},\hat{\rho}^\prime)}_bX^{A0(\rho,\hat{\rho}^\prime)}~,\\
X^{0}_A\hat{t}_aX^{A0}t_b=&\sum_{\rho,\rho^\prime,\hat{\rho},\hat{\rho}^{\prime}}X^{0(\rho,\hat{\rho})}_A\hat{\sigma}^{(\hat{\rho},\hat{\rho}^\prime)}_aX^{A0(\rho^\prime,\hat{\rho}^\prime)}\sigma^{(\rho^\prime,\rho)}_b~,\\
}
}
where $\sigma^{(\rho,\rho^\prime)}_a\equiv\bra{\rho}t_a\ket{\rho^\prime}$ and we used the fact that $\sum_\rho\ket{\rho}\bra{\rho}=\mathds{1}$. 
We then define the following matrices:
\eq{
\spl{
&\mathds{B}_{ab}=X^{A0}t_{(a}t_{b)}X^0_A~,\\
&\mathds{C}_{ab}=X^{0}_A\hat{t}_{(a}\hat{t}_{b)}X^{A0}~,\\
&\mathds{D}_{ab}=-X^{0}_A\hat{t}_aX^{A0}t_b~,
}
}
and the deformations that are quadratic in gauge fields can be represented as:
\eq{
\label{mat}
\begin{array}{cccc}
&\begin{array}{cc}
A&\ \ \ \ \ \ \ \ \ \ \ \ \ \ \ \ \ \hat{A}
\end{array}&\\
\begin{array}{c}
A\\
\hat{A}
\end{array}&
\begin{pmatrix}
\mathds{B}+4\pi^2(\vec{n}\cdot\vec{n})\times\mathds{1}&\mathds{D}^{Tr}\\
\mathds{D}&\mathds{C}+4\pi^2(\vec{n}\cdot\vec{n})\times\mathds{1}
\end{pmatrix}&\otimes\ \Gamma~.
\end{array}
}
The determinant of 
the tensor product of two matrices $A$ and $B$ is given by:
\eq{
\det(A\otimes B)=(\det A)^{\dim B} (\det B)^{\dim A}~.
} 
Therefore, when $n_z\neq0$, we have:
\eq{
\spl{
\det(A,\hat{A})|_{n_z\neq0}=&\prod_{(\vec{n},-\vec{n}),n_z\neq0}\{(\det\mathds{A})^4\times(\prod_a\det\Gamma)^2\}\\
=&\prod_{(\vec{n},-\vec{n}),n_z\neq0}\{(\det\mathds{A})^4\times(\prod_a\frac{(\vec{n}\cdot\vec{n})^2}{n^4_z})^2\}\\
=&\prod_{(\vec{n},-\vec{n}),n_z\neq0}\{(\det[\frac{\mathds{A}}{4\pi^2({\vec{n}\cdot\vec{n}})}])^4\times(\prod_a\frac{(4\pi^2)^4(\vec{n}\cdot\vec{n})^6}{n^4_z})^2\}\\
=&\prod_{\vec{n},n_z\neq0}\{(\det[\frac{\mathds{A}}{4\pi^2({\vec{n}\cdot\vec{n}})}])^2\times(\prod_a\frac{16\pi^4(\vec{n}\cdot\vec{n})^3}{n^2_z})^2\}~,
}
}
where:
\eq{
\mathds{A}\equiv
\begin{pmatrix}
\mathds{B}+4\pi^2(\vec{n}\cdot\vec{n})\times\mathds{1}&\mathds{D}^{Tr}\\
\mathds{D}&\mathds{C}+4\pi^2(\vec{n}\cdot\vec{n})\times\mathds{1}
\end{pmatrix}~.
}
For the case where $n_z=0$, but $n_x$ or $n_y$ are not equal to zero, the procedure is similar. The determinant coming from integrating over $A_\mu$ reads:
\eq{
\spl{
\det(A,\hat{A})=&\prod_{\vec{n}}(\det(\frac{\mathds{A}}{4\pi^2({\vec{n}\cdot\vec{n}})}))^2\prod_a\{\prod_{\vec{n},n_z\neq0}[16\pi^4\frac{(\vec{n}\cdot\vec{n})^3}{n^2_z}]^2\\
&\times\prod_{\vec{n},n_z=0,n_x\neq0} [16\pi^4\frac{(\vec{n}\cdot\vec{n})^3}{n^2_x}]^2\prod_{\vec{n},n_z=n_x=0,n_y\neq0}[16\pi^4\frac{(\vec{n}\cdot\vec{n})^3}{n^2_y}]^2\}~.
}
}
The contribution to the one-loop determinant coming from the terms involving gauge fields is thus:
\eq{
\label{r14}
\spl{
Z_{1-loop}(A,\hat{A})=&\frac{\prod_a\{\prod_{\vec{n},n_z\neq0}n^2_z\prod_{\vec{n},n_z=0,n_x\neq0}n^2_x\prod_{\vec{n},n_z=n_x=0,n_y\neq0}n^2_y\}}{\prod_a\prod_{\vec{n}}16\pi^4(\vec{n}\cdot\vec{n})^3}\\
&\times\prod_{\vec{n}}(\det[\frac{\mathds{A}}{4\pi^2({\vec{n}\cdot\vec{n}})}])^{-1}~.
}
}
One may worry about regularizing the numerator. However, we note that the gauge-fixing delta function also gives a Jacobian factor to the one-loop determinant. Indeed in the ghost action we have:
\eq{
\spl{
&\exp\{i\int d^3x\mathrm{tr}(b\partial^\mu A_\mu)\}\\
=&\exp\{i2\pi\sum_{\vec{n}}\sum_ab^a_{-\vec{n}}(\vec{n}\cdot \vec{A}^a_{\vec{n}})\}~.
}
}
After integrating out $b^a_{\vec{n}}$ we obtain:
\eq{
\prod_{\vec{n}}\prod_a\delta(\vec{n}\cdot \vec{A}^a_{\vec{n}})~.
}
This product of delta functions imposes the gauge-fixing Lorentz condition and, upon integrating out $A_\mu$, $\hat{A}_\mu$, gives a Jacobian factor which cancels the numerator of (\ref{r14}).

The integral over $F_A$ simply contributes an overall constant factor. Finally we are left with the integration over $X^\prime_A$:
\eq{
\spl{
&\int d^3x\mathrm{tr}\{-X^\prime_A\DAlambert X^{\prime A}\}\\
=&\sum_{\vec{n}}4\pi^2\vec{n}^2\mathrm{tr}\{X^\prime_{A,-\vec{n}}X^{\prime A}_{\vec{n}}\}\\
=&\sum_{\vec{n}}2\pi^2\vec{n}^2\mathrm{tr}\{X^\prime_{A,-\vec{n}}X^{\prime A}_{\vec{n}}+X^{\prime A}_{\vec{n}}X^\prime_{A,-\vec{n}}\}\\
=&\sum_{(\vec{n},-\vec{n})}2\pi^2\vec{n}^2\mathrm{tr}\{X^\prime_{A,-\vec{n}}X^{\prime A}_{\vec{n}}+X^\prime_{A,\vec{n}}X^{\prime A}_{-\vec{n}}\\
&+X^{\prime A}_{\vec{n}}X^\prime_{A,-\vec{n}}+X^{\prime A}_{-\vec{n}}X^\prime_{A,\vec{n}}\}~.
}
}
This integration is Gaussian, and the corresponding determinant is:
\eq{
\det{X^\prime_A}=\prod_A\prod_{(\rho,\hat{\rho})}\prod_{\vec{n}}(2\pi^2\vec{n}^2)^2~,
}
where $(\rho,\hat{\rho})$ runs over the weights of the bifundamental representation. 
Therefore, the total  contribution of the bosonic part to the one-loop determinant reads:
\eq{
\label{r17}
\spl{
Z_{1-loop}(Boson)=&\frac{1}{\{\prod_a\prod_{\vec{n}}16\pi^4(\vec{n}^2)^3\}|_{A,\hat{A}}\{\prod_A\prod_{(\rho,\hat{\rho})}\prod_{\vec{n}}2\pi^2\vec{n}^2\}|_{X^\prime}}\\
&\times\prod_{\vec{n}}(\det[\frac{\mathds{A}}{4\pi^2({\vec{n}\cdot\vec{n}})}])^{-1}\\
=&\frac{1}{\{\prod_{\vec{n}}[16\pi^4(\vec{n}^2)^3]^d\}\{\prod_A\prod_{\vec{n}}(2\pi^2\vec{n}^2)^{w^2}\}}\\
&\times\prod_{\vec{n}}(\det[\frac{\mathds{A}}{4\pi^2({\vec{n}\cdot\vec{n}})}])^{-1}~.
}
}
Here $d$ is the dimension of the gauge group and $w$ is the dimension of its fundamental representation. For $U(N)$ in particular we have $d=N^2$, $w=N$.

\subsubsection{Determinant from Fermions}

The fermionic part of the deformation is:
\eq{
\spl{
&\int dx^3\mathrm{tr}\{-i\tilde{\lambda}\slashed{\partial}\lambda\}+\int dx^3\mathrm{tr}\{-i\tilde{\hat{\lambda}}\slashed{\partial}\hat{\lambda}\}+\int dx^3\mathrm{tr}\{-i\tilde{\Psi}^A\slashed{\partial}\Psi_A\\
&+\Omega^{AB}\tilde{\lambda}X^0_B\Psi_A+\Omega_{AB}\tilde{\hat{\lambda}}X^{0B}\Psi^A-\Omega^{AB}X^0_B\tilde{\hat{\lambda}}\Psi_A-\Omega_{AB}X^{0B}\tilde{\lambda}\Psi^A\}~.
}
}
Using the expansion $\lambda=\lambda_+\epsilon+\lambda_-\epsilon^c$ for the gaugino kinetic term, cf. appendix C, we have:
\eq{
\spl{
&\int dx^3\mathrm{tr}\{-i\tilde{\lambda}\slashed{\partial}\lambda\}\\
=&\int dx^3\mathrm{tr}\{-i(\lambda_+V\cdot\partial\lambda_+-\lambda_-\bar{V}\cdot\partial\lambda_--\lambda_-U\cdot\partial\lambda_+-\lambda_+U\cdot\partial\lambda_-)\}\\
=&2\pi\sum_a\sum_{\vec{n}}\{V\cdot\vec{n}\lambda^a_{+,-\vec{n}}\lambda^a_{+,\vec{n}}-\bar{V}\cdot\vec{n}\lambda^a_{-,-\vec{n}}\lambda^a_{-,\vec{n}}-U\cdot\vec{n}\lambda^a_{-,-\vec{n}}\lambda^a_{+,\vec{n}}\\
&-U\cdot\vec{n}\lambda^a_{+,-\vec{n}}\lambda^a_{-,\vec{n}}\}\\
=&2\pi\sum_a\sum_{(\vec{n},-\vec{n})}\{(V\cdot\vec{n}\lambda^a_{+,-\vec{n}}\lambda^a_{+,\vec{n}}-\bar{V}\cdot\vec{n}\lambda^a_{-,-\vec{n}}\lambda^a_{-,\vec{n}}-U\cdot\vec{n}\lambda^a_{-,-\vec{n}}\lambda^a_{+,\vec{n}}\\
&-U\cdot\vec{n}\lambda^a_{+,-\vec{n}}\lambda^a_{-,\vec{n}})+(-V\cdot\vec{n}\lambda^a_{+,\vec{n}}\lambda^a_{+,-\vec{n}}+\bar{V}\cdot\vec{n}\lambda^a_{-,\vec{n}}\lambda^a_{-,-\vec{n}}\\
&+U\cdot\vec{n}\lambda^a_{-,\vec{n}}\lambda^a_{+,-\vec{n}}+U\cdot\vec{n}\lambda^a_{+,\vec{n}}\lambda^a_{-,-\vec{n}})\}~,
}
}
where we symmetrized the indices $+$, $-$ and $\vec{n}$, $-\vec{n}$ of the gaugini in the last equation. For each pair of $(\vec{n},-\vec{n})$ and each $a$, this 
can be written in matrix notation as:
\eq{
\begin{array}{cc}
&\begin{array}{cccc}
\lambda^a_{+,\vec{n}}&\ \ \ \ \ \lambda^a_{-,\vec{n}}&\ \ \ \ \ \lambda^a_{+,-\vec{n}}&\ \ \ \ \ \lambda^a_{-,-\vec{n}}
\end{array}\\
\begin{array}{c}
\lambda^a_{+,\vec{n}}\\
\lambda^a_{-,\vec{n}}\\
\ \lambda^a_{+,-\vec{n}}\\
\ \lambda^a_{-,-\vec{n}}
\end{array}&
\begin{pmatrix}
0&0&-2\pi V\cdot\vec{n}&2\pi U\cdot\vec{n}\\
0&0&2\pi U\cdot\vec{n}&2\pi \bar{V}\cdot\vec{n}\\
2\pi V\cdot\vec{n}&-2\pi U\cdot\vec{n}&0&0\\
-2\pi U\cdot\vec{n}&-2\pi \bar{V}\cdot\vec{n}&0&0
\end{pmatrix}~.
\end{array}
}
Similarly for the matter fermion kinetic term:
\eq{
\spl{
&2\pi\sum_{(\vec{n},-\vec{n})}\mathrm{tr}\{(V\cdot\vec{n}\Psi^A_{+,-\vec{n}}\Psi_{A+,\vec{n}}-\bar{V}\cdot\vec{n}\Psi^A_{-,-\vec{n}}\Psi_{A-,\vec{n}}-U\cdot\vec{n}\Psi^A_{-,-\vec{n}}\Psi_{A+,\vec{n}}\\
&-U\cdot\vec{n}\Psi^A_{+,-\vec{n}}\Psi_{A-,\vec{n}})+(-V\cdot\vec{n}\Psi^A_{+,\vec{n}}\Psi_{A+,-\vec{n}}+\bar{V}\cdot\vec{n}\Psi^A_{-,\vec{n}}\Psi_{A-,-\vec{n}}\\
&+U\cdot\vec{n}\Psi^A_{-,\vec{n}}\Psi_{A+,-\vec{n}}+U\cdot\vec{n}\Psi^A_{+,\vec{n}}\Psi_{A-,-\vec{n}})\}\\
=&\pi\sum_{(\vec{n},-\vec{n})}\mathrm{tr}\{(V\cdot\vec{n}\Psi^A_{+,-\vec{n}}\Psi_{A+,\vec{n}}-\bar{V}\cdot\vec{n}\Psi^A_{-,-\vec{n}}\Psi_{A-,\vec{n}}-U\cdot\vec{n}\Psi^A_{-,-\vec{n}}\Psi_{A+,\vec{n}}\\
&-U\cdot\vec{n}\Psi^A_{+,-\vec{n}}\Psi_{A-,\vec{n}})+(-V\cdot\vec{n}\Psi^A_{+,\vec{n}}\Psi_{A+,-\vec{n}}+\bar{V}\cdot\vec{n}\Psi^A_{-,\vec{n}}\Psi_{A-,-\vec{n}}\\
&+U\cdot\vec{n}\Psi^A_{-,\vec{n}}\Psi_{A+,-\vec{n}}+U\cdot\vec{n}\Psi^A_{+,\vec{n}}\Psi_{A-,-\vec{n}})\}+(-1)\Psi^A\leftrightarrow\Psi_A~.
}
}
The last term arises due to the symmetrization of $\Psi^A$ and $\Psi_A$. When decomposed into the weight spaces, this becomes:
\eq{
\spl{
&\pi\sum_{(\rho,\hat{\rho})}\sum_{(\vec{n},-\vec{n})}\{(V\cdot\vec{n}\Psi^{A(\rho,\hat{\rho})}_{+,-\vec{n}}\Psi^{(\rho,\hat{\rho})}_{A+,\vec{n}}-\bar{V}\cdot\vec{n}\Psi^{A(\rho,\hat{\rho})}_{-,-\vec{n}}\Psi^{(\rho,\hat{\rho})}_{A-,\vec{n}}-U\cdot\vec{n}\Psi^{A(\rho,\hat{\rho})}_{-,-\vec{n}}\Psi^{(\rho,\hat{\rho})}_{A+,\vec{n}}\\
&-U\cdot\vec{n}\Psi^{A(\rho,\hat{\rho})}_{+,-\vec{n}}\Psi^{(\rho,\hat{\rho})}_{A-,\vec{n}})+(-V\cdot\vec{n}\Psi^{A(\rho,\hat{\rho})}_{+,\vec{n}}\Psi^{(\rho,\hat{\rho})}_{A+,-\vec{n}}+\bar{V}\cdot\vec{n}\Psi^{A(\rho,\hat{\rho})}_{-,\vec{n}}\Psi^{(\rho,\hat{\rho})}_{A-,-\vec{n}}\\
&+U\cdot\vec{n}\Psi^{A(\rho,\hat{\rho})}_{-,\vec{n}}\Psi^{(\rho,\hat{\rho})}_{A+,-\vec{n}}+U\cdot\vec{n}\Psi^{A(\rho,\hat{\rho})}_{+,\vec{n}}\Psi^{(\rho,\hat{\rho})}_{A-,-\vec{n}})\}+(-1)\Psi^A\leftrightarrow\Psi_A~.
}
}
For each pair of weights $(\rho,\hat{\rho})$ and each pair of $(\vec{n},-\vec{n})$,  these terms can be written with the help of two matrices:
\eq{
\begin{array}{cc}
&\begin{array}{cccc}
\Psi^{(\rho,\hat{\rho})}_{A+,\vec{n}}&\ \Psi^{(\rho,\hat{\rho})}_{A-,\vec{n}}&\ \Psi^{(\rho,\hat{\rho})}_{A+,-\vec{n}}&\ \Psi^{(\rho,\hat{\rho})}_{A-,-\vec{n}}
\end{array}\\
\begin{array}{c}
\Psi^{A(\rho,\hat{\rho})}_{+,\vec{n}}\\
\Psi^{A(\rho,\hat{\rho})}_{-,\vec{n}}\\
\Psi^{A(\rho,\hat{\rho})}_{+,-\vec{n}}\\
\Psi^{A(\rho,\hat{\rho})}_{-,-\vec{n}}
\end{array}&
\begin{pmatrix}
0&0&-\pi V\cdot\vec{n}&\pi U\cdot\vec{n}\\
0&0&\pi U\cdot\vec{n}&\pi\bar{V}\cdot\vec{n}\\
\pi V\cdot\vec{n}&-\pi U\cdot\vec{n}&0&0\\
-\pi U\cdot\vec{n}&-\pi\bar{V}\cdot\vec{n}&0&0\\
\end{pmatrix}~,
\end{array}
}
and:
\eq{
\begin{array}{cc}
&\begin{array}{cccc}
\Psi^{A(\rho,\hat{\rho})}_{+,\vec{n}}&\ \Psi^{A(\rho,\hat{\rho})}_{-,\vec{n}}&\ \Psi^{A(\rho,\hat{\rho})}_{+,-\vec{n}}&\ \Psi^{A(\rho,\hat{\rho})}_{-,-\vec{n}}
\end{array}\\
\begin{array}{c}
\Psi^{(\rho,\hat{\rho})}_{A+,\vec{n}}\\
\Psi^{(\rho,\hat{\rho})}_{A-,\vec{n}}\\
\ \ \Psi^{(\rho,\hat{\rho})}_{A+,-\vec{n}}\\
\ \ \Psi^{(\rho,\hat{\rho})}_{A-,-\vec{n}}
\end{array}&
\begin{pmatrix}
0&0&-\pi V\cdot\vec{n}&\pi U\cdot\vec{n}\\
0&0&-\pi V\cdot\vec{n}&\pi U\cdot\vec{n}\\
\pi V\cdot\vec{n}&-\pi U\cdot\vec{n}&0&0\\
-\pi U\cdot\vec{n}&-\pi\bar{V}\cdot\vec{n}&0&0
\end{pmatrix}~.
\end{array}
}
Similarly, the Yukawa interactions can be written as:
\eq{
\spl{
&\int dx^3\mathrm{tr}\{\Omega^{AB}\tilde{\lambda}X^0_B\Psi_A+\Omega_{AB}\tilde{\hat{\lambda}}X^{0B}\Psi^A-\Omega^{AB}X^0_B\tilde{\hat{\lambda}}\Psi_A-\Omega_{AB}X^{0B}\tilde{\lambda}\Psi^A\}\\
&=a\sum_{\vec{n}}\mathrm{tr}\{(\Psi_{A+,-\vec{n}}\lambda_{-,\vec{n}}-\Psi_{A-,-\vec{n}}\lambda_{+,\vec{n}})\Omega^{AB}X^0_B+(\Psi^A_{+,-\vec{n}}\hat{\lambda}_{-,\vec{n}}-\Psi^A_{-,-\vec{n}}\hat{\lambda}_{+,\vec{n}})\Omega_{AB}X^{0B}\\
&-\Omega^{AB}X^0_B(\hat{\lambda}_{+,-\vec{n}}\Psi_{A-,\vec{n}}-\hat{\lambda}_{-,-\vec{n}}\Psi_{A+,\vec{n}})-\Omega_{AB}X^{0B}(\lambda_{+,-\vec{n}}\Psi^A_{-,\vec{n}}-\lambda_{-,-\vec{n}}\Psi^A_{+,\vec{n}})\}\\
&=a\sum_{(\vec{n},-{\vec{n}})}\mathrm{tr}\{(\Psi_{A+,-\vec{n}}\lambda_{-,\vec{n}}-\Psi_{A-,-\vec{n}}\lambda_{+,\vec{n}})\Omega^{AB}X^0_B+(\Psi^A_{+,-\vec{n}}\hat{\lambda}_{-,\vec{n}}-\Psi^A_{-,-\vec{n}}\hat{\lambda}_{+,\vec{n}})\Omega_{AB}X^{0B}\\
&-\Omega^{AB}X^0_B(\hat{\lambda}_{+,-\vec{n}}\Psi_{A-,\vec{n}}-\hat{\lambda}_{-,-\vec{n}}\Psi_{A+,\vec{n}})-\Omega_{AB}X^{0B}(\lambda_{+,-\vec{n}}\Psi^A_{-,\vec{n}}-\lambda_{-,-\vec{n}}\Psi^A_{+,\vec{n}})\\
&+(\Psi_{A+,\vec{n}}\lambda_{-,-\vec{n}}-\Psi_{A-,\vec{n}}\lambda_{+,-\vec{n}})\Omega^{AB}X^0_B+(\Psi^A_{+,\vec{n}}\hat{\lambda}_{-,-\vec{n}}-\Psi^A_{-,\vec{n}}\hat{\lambda}_{+,-\vec{n}})\Omega_{AB}X^{0B}\\
&-\Omega^{AB}X^0_B(\hat{\lambda}_{+,\vec{n}}\Psi_{A-,-\vec{n}}-\hat{\lambda}_{-,\vec{n}}\Psi_{A+,-\vec{n}})-\Omega_{AB}X^{0B}(\lambda_{+,\vec{n}}\Psi^A_{-,-\vec{n}}-\lambda_{-,\vec{n}}\Psi^A_{+,-\vec{n}})\}~.
}
}
Each term, such as  $\mathrm{tr}\{\Psi_{A+,-\vec{n}}\lambda_{-,\vec{n}}\Omega^{AB}X^0_B\}$ for example, can be written in terms of the algebra representations as follows:
\eq{
\spl{
&\sum_{(\rho,\hat{\rho})}\sum_{(\rho^\prime,\hat{\rho}^\prime)}\sum_a\Psi^{(\rho,\hat{\rho})}_{A+,-\vec{n}}\bra{\hat{\rho}}\bra{\rho}\lambda^a_{-,\vec{n}}t_a\ket{\rho^\prime}\ket{\hat{\rho}^\prime}\Omega^{AB}X^{0(\rho^\prime,\hat{\rho}^\prime)}_B\\
=&\sum_{\rho,\rho^\prime,\hat{\rho}}\sum_a\Psi^{(\rho,\hat{\rho})}_{A+,-\vec{n}}\lambda^a_{-,\vec{n}}\sigma^{(\rho,\rho^\prime)}_a\Omega^{AB}X^{0(\rho^\prime,\hat{\rho})}_B~,
}
}
where $\sigma^{(\rho,\rho^\prime)}_a\equiv\bra{\rho}t_a\ket{\rho^\prime}$ ($\hat{\sigma}^{(\hat{\rho},\hat{\rho}^\prime)}_a\equiv\bra{\hat{\rho}}\hat{t}_a\ket{\hat{\rho}^\prime}$). 
Therefore the matrix elements for each $\Psi^{(\rho,\hat{\rho})}_A$ and each $\lambda^a$ are:
\eq{
\begin{array}{cc}
&\begin{array}{cccc}
\lambda^a_{+,\vec{n}}&\ \lambda^a_{-,\vec{n}}&\ \lambda^a_{+,-\vec{n}}&\ \lambda^a_{-,-\vec{n}}
\end{array}\\
\begin{array}{c}
\Psi^{(\rho,\hat{\rho})}_{A+,\vec{n}}\\
\Psi^{(\rho,\hat{\rho})}_{A-,\vec{n}}\\
\ \ \Psi^{(\rho,\hat{\rho})}_{A+,-\vec{n}}\\
\ \ \Psi^{(\rho,\hat{\rho})}_{A-,-\vec{n}}
\end{array}&
\begin{pmatrix}
0&0&0&[\sigma X]\\
0&0&-[\sigma X]&0\\
0&[\sigma X]&0&0\\
-[\sigma X]&0&0&0
\end{pmatrix}~,
\end{array}~~
\begin{array}{cc}
&\begin{array}{cccc}
\ \Psi^{(\rho,\hat{\rho})}_{A+,\vec{n}}&\Psi^{(\rho,\hat{\rho})}_{A-,\vec{n}}&\Psi^{(\rho,\hat{\rho})}_{A-,\vec{n}}&\Psi^{(\rho,\hat{\rho})}_{A-,-\vec{n}}
\end{array}\\
\begin{array}{c}
\lambda^a_{+,\vec{n}}\\
\lambda^a_{-,\vec{n}}\\
\ \lambda^a_{+,-\vec{n}}\\
\ \lambda^a_{-,-\vec{n}}
\end{array}
&\begin{pmatrix}
0&0&0&[\sigma X]\\
0&0&-[\sigma X]&0\\
0&[\sigma X]&0&0\\
-[\sigma X]&0&0&0
\end{pmatrix}~,
\end{array}
}
where $[\sigma X]\equiv\frac{1}{2}\sigma^{(\rho,\rho^\prime)}_a\Omega^{AB}X^{0(\rho^\prime,\hat{\rho})}_B$ and $\lambda$, $\Psi$ are symmetrized. This explains the factor $\frac{1}{2}$ in each entry. A summation over $\rho^\prime$ is understood in $\sigma^{(\rho,\rho^\prime)}_a\Omega^{AB}X^{0(\rho^\prime,\hat{\rho})}_B$.

The fermionic part of the deformation for each pair of $(\vec{n},-\vec{n})$ can be written in matrix notation as:
\eq{
\label{r15}
\begin{array}{cc}
&\begin{array}{cccc}
\ \ \ \lambda^a&\ \ \ \ \ \ \hat{\lambda}^{a^\prime}&\ \ \ \ \Psi^{A(\rho,\hat{\rho})}&\ \  \ \ \Psi^{(\rho,\hat{\rho})}_A
\end{array}\\
\begin{array}{c}
\lambda^a\\
\hat{\lambda}^{a^\prime}\\
\Psi^{A(\rho,\hat{\rho})}\\
\Psi^{(\rho,\hat{\rho})}_A\\
\end{array}
&\begin{pmatrix}
M&0&(X\sigma)_A&-(\sigma X)^A\\
0&M&-\widehat{(\sigma X)}_A&\widehat{(X\sigma)}^A\\
(X\sigma)_A&-\widehat{(\sigma X)}_A&0&N\\
-(\sigma X)^A&\widehat{(X\sigma)}^A&N&0
\end{pmatrix}~,
\end{array}
}
where:
\eq{
\spl{
\label{def1}
&M=2N\equiv\begin{pmatrix}
0&0&-2\pi V\cdot\vec{n}&2\pi U\cdot\vec{n}\\
0&0&2\pi U\cdot\vec{n}&2\pi\bar{V}\cdot\vec{n}\\
2\pi V\cdot\vec{n}&-2\pi U\cdot\vec{n}&0&0\\
-2\pi U\cdot\vec{n}&-2\pi\bar{V}\cdot\vec{n}&0&0
\end{pmatrix}~,\\
&(\sigma X)^A\equiv\frac{1}{2}\sigma^{(\rho,\rho^\prime)}_a\Omega^{AB}X^{0(\rho^\prime,\hat{\rho})}_B\times\mathds{S}~,\\
&(X\sigma)_A\equiv\frac{1}{2}\Omega_{AB}X^{B0(\rho^\prime,\hat{\rho})}\sigma^{(\rho^\prime,\rho)}_a\times\mathds{S}~,\\
&\widehat{(\sigma X)}_A\equiv\frac{1}{2}\hat{\sigma}^{(\hat\rho,{\hat\rho}^\prime)}_a\Omega_{AB}X^{B0(\rho,\hat{\rho}^\prime)}\times\mathds{S}~,\\
&\widehat{(X\sigma)}^A\equiv\frac{1}{2}\Omega^{AB}X^{0(\rho,\hat{\rho}^\prime)}_B\hat{\sigma}^{(\hat{\rho}^\prime,\hat{\rho})}_a\times\mathds{S}~,
}
}
and:
\eq{
\label{def2}
\mathds{S}\equiv\begin{pmatrix}0&0&0&-1\\0&0&1&0\\0&-1&0&0\\1&0&0&0\end{pmatrix}~.
}
As before we have $a,a'=1,\dots,d$; $\rho,\hat{\rho}=1,\dots,w$; $A=1,\dots,4$, 
where $d$ is the dimension of the gauge group and $w$ is the dimension of its fundamental representation. Therefore (\ref{r15}) is a $2d+8w^2$ by $2d+8w^2$ block matrix: each entry is given by one of the above four by four matrices.

The matrix (\ref{r15}) can be partitioned into four blocks:
\eq{
\label{r21}
\begin{pmatrix}
A_{8d\times8d}&B_{8d\times32w^2}\\
C_{32w^2\times8d}&D_{32w^2\times32w^2}
\end{pmatrix}
:=
\left(
\begin{array}{cc;{2pt/2pt}cc}
M&0&(X\sigma)_A&-(\sigma X)^A\\
0&M&-\widehat{(\sigma X)}_A&\widehat{(X\sigma)}^A\\
\hdashline[2pt/2pt]
(X\sigma)_A&-\widehat{(\sigma X)}_A&0&N\\
-(\sigma X)^A&\widehat{(X\sigma)}^A&N&0
\end{array}\right)
~,
}
so that the determinant reads:\footnote{We use the notation $\mathds{A},\mathds{B},\mathds{C},\mathds{D}$ for the matrices in the bosonic  sector, while the matrices $A,B,C,D$ are used for the fermion fields. We hope this does not cause any confusion with the $Sp(2)$ indices.}
\eq{
\det\begin{pmatrix}
A&B\\
C&D
\end{pmatrix}
=\det A\det D\det[\mathds{1}-D^{-1}CA^{-1}B]~.
}
The determinants 
$\det A$ and $\det D$ are straightforward to compute:
\al{
&\det A=(\det M)^{2d}=[16\pi^4(\vec{n}^2)^2]^{2d}~,\\
&\det D=\prod_A(\det N)^{2w^2}=\prod_A[\pi^4(\vec{n}^2)^2]^{2w^2}~.
}
Their combined contribution to the one-loop determinant is:
\eq{
\label{r16}
\prod_{\vec{n}}\{(4\pi^2\vec{n}^2)^d\prod_A(\pi^2\vec{n}^2)^{w^2}\}~.
}
Furthermore the integrations over the ghosts and anti-ghosts for the two gauge groups contribute $(\det\DAlambert)^2=\{\prod_{\vec{n}}(4\pi^2\vec{n}^2)^d\}^2$. When combined with (\ref{r16}) this gives:
\eq{
\prod_{\vec{n}}\{(4\pi^2\vec{n}^2)^{3d}\prod_A(\pi^2\vec{n}^2)^{w^2}\}~.
}
Up to a constant factor, this partially cancels the one-loop determinant from the boson sector, (\ref{r17}). We are thus left with only $X^0$-dependent contributions from both boson and fermion sectors.

Inserting the localization conditions (\ref{r22}) into the off-shell Lagrangian (\ref{r19}) gives a vanishing classical contribution. Therefore the partition function is given purely by the one-loop determinant:
\eq{
Z=\int \prod_A\prod_{(\rho,\hat{\rho})}dX^{0(\rho,\hat{\rho})}_A \prod_B\prod_{(\rho^\prime,\hat{\rho}^\prime)}dX^{B0(\rho^\prime,\hat{\rho}^\prime)}\frac{\prod_{(\vec{n},-\vec{n})}\{\det[\mathds{1}-D^{-1}CA^{-1}B]\}^{\frac{1}{2}}}{\prod_{\vec{n}}\det[\frac{\mathds{A}}{4\pi^2({\vec{n}\cdot\vec{n}})}]}~.
}
We now make use of the Sylvester identity:
\eq{
\det[\mathds{1}-D^{-1}CA^{-1}B]=\det[\mathds{1}-BD^{-1}CA^{-1}]~,
}
where the matrix on the left-hand side above is $32w^2\times32w^2$, while 
the matrix on the right-hand side is $8d\times8d$. 
Using the definitions in (\ref{def1}) and (\ref{def2}), one can show that:
\eq{
\spl{
&\det[\mathds{1}-BD^{-1}CA^{-1}]\\
=&\det[\mathds{1}+C^{Tr}D^{-1}CA^{-1}]\\
=&\det[\mathds{1}+\begin{pmatrix}
\mathds{B}&\mathds{D}^{Tr}\\
\mathds{D}&\mathds{C}
\end{pmatrix}\otimes\frac{\mathds{S}N^{-1}\mathds{S}M^{-1}}{2}]\\
=&\det[\mathds{1}+\begin{pmatrix}
\mathds{B}&\mathds{D}^{Tr}\\
\mathds{D}&\mathds{C}
\end{pmatrix}\otimes\frac{\mathds{1}_{4\times4}}{4\pi^2(\vec{n}\cdot\vec{n})}]\\
=&\{\det[\frac{\mathds{A}}{4\pi^2({\vec{n}\cdot\vec{n}})}]\}^4~.
}
}
Putting this back into the one-loop determinant, we see that the fermion and boson determinants cancel exactly against each other.

\vfill\break

\section{Discussion}

We have partially carried out the localization procedure for the $\mathcal{N}=1$ Chern-Simons-matter theory on $T^3$ with periodic boundary conditions. In particular we computed the contributions to the partition function from the locus of saddle points with vanishing gauge connection. As expected, restricting to this locus gives a trivial contribution to the partition function, i.e.~the bosonic and fermionic contributions exactly cancel each other. Indeed evaluating
the partition function on the 
flat torus at the trivial vacuum (vanishing gauge connection) 
simply counts the degrees of freedom of the theory, and for a supersymmetric theory one expects a complete cancellation. Of course the full partition function should receive contributions also from saddle points 
with nonvanishing flat gauge connections, which we have not computed here. 
We hope to return to this in the future. 

Another potentially interesting direction in which this paper may be generalized is by allowing 
for a more general Killing spinor equation than the eq.~(\ref{Killing}) which was used for the present analysis. This may be achieved by coupling to a supergravity background  and could 
provide additional possibilities for spaces on which the theory localizes.

The authors of \cite{Knodel:2014xea} considered Euclidean 4d $\mathcal{N}=1$ theories without R-symmetry, and concluded that no localization 
is possible in this case. Our results are not in contradiction with their conclusions. Indeed it is possible to construct 3d theories without R-symmetry by dimensional reduction and further truncation of 4d theories {\it with} R-symmetry.

Our results 
have the following implication for 
the partition function of the ABJM model on $T^3$.\footnote{Note that placing  
the ABJM model on $T^3$ breaks conformal invariance: this can be seen directly from the fact that the superconformal transfromation parameter $\eta$ of section 
\ref{sec:conf} is not well-defined on the torus, being linear in the coordinates.}  
Our analysis of the saddle points shows that the classical CS action vanishes on the locus of flat gauge connections on $T^3$, cf.~(\ref{s}). Since the one-loop determinant around the saddle points does not introduce any dependence on the two CS levels, it follows by the localization argument that the partition function is independent of the level $k\equiv k_1=-k_2$. Hence we may compute the partition function in the limit $k\rightarrow\infty$ with $N$ fixed, which corresponds to vanishing 't Hooft coupling. In this limit the matter sector becomes free and decouples from the CS action. Therefore the resulting partition function factorizes into a pure supersymmetric CS partition function and a free matter piece. The latter is trivial, i.e.~the bosonic and fermionic contributions exactly cancel each other. Moreover our localization results can be applied to the pure CS partition function to show that the contribution from the saddle points with vanishing gauge connection is also trivial. As
mentioned above, this is consistent with what one expects for a supersymmetric theory.

\acknowledgments

We would like to thank Ergin Sezgin, Alessandro Tomasiello and especially Yi Pang for useful discussions. Y.~Z. would also like to thank 
J.~K\"{a}ll\'{e}n, K.~Ohta, V.~Pestun, D.~Robbins and A.~Royston. Y.~Z. was financially supported by the French governement Chateaubriand Fellowships Programme 2015 and by the Mitchell Institute for Fundamental Physics and Astronomy. Y.~Z. would also like to acknowledge the hospitality of the Institut de Physique Nucl\'{e}aire de Lyon, where most of the present work was done.


\appendix

\section{Spinor and gamma-matrix conventions in 3d}\label{app:sc}

The charge conjugation matrix in three dimensions satisfies:
\eq{\label{a4}
C^{Tr}=-C; ~~~~~~ (C\g^\mu)^{Tr}=C\g^\mu; ~~~~~~ C^*=-C^{-1}~.
}
For any spinor $\psi$ and in any spacetime signature we define:
\eq{\widetilde{\psi}\equiv\psi^{Tr}C^{-1}
~.}
Moreover in Euclidean signature we define:
\eq{\psi^c\equiv C\psi^*
~.}
It follows that,
\eq{\label{a6}\psi^{\dagger}=-\tilde{\psi^c}~;
~~~(\psi^c)^c=-\psi
~.
}
The irreducible spinor representation in three Euclidean dimensions is two-dimensional complex (pseudoreal).

The Gamma matrices in Euclidean signature are taken to obey:
\eq{
(\g_{\mu})^{\dagger}=\g_{\mu}~.
}
Antisymmetric products of Gamma matrices are defined by:
\eq{
\g^{(n)}_{{\mu}_1\dots \mu_n}\equiv\g_{[\mu_1}\dots\g_{\mu_n]}~.
}
In Euclidean signature the Hodge-dual of an antisymmetric product of
gamma matrices is given by:
\eq{
\star\g_{(n)}={{(-1)^{\frac{1}{2}n(n-1)}}}\g_{(3-n)}~.
\label{hodge2}
}

\section{$\mathcal{N}=1$ Superconformal Symmetry}

\subsection{Poincar\'e Supersymmetry}

In this subsection we show the invariance of the on-shell Lagrangian (\ref{r20}) under the Poincar\'e supersymmetry.

The most general Poincar\'e supersymmetry transformations read:
\begin{equation}
\begin{split}
\delta X_A&=i\Omega_{AB}\tilde{\epsilon}\Psi^B \\
\delta X^A&=i\Omega^{AB}\tilde{\epsilon}\Psi_B \\
\delta\Psi_A&=\Omega_{AB}\gamma^{\mu}\epsilon D_{\mu}X^B+\delta_3\Psi_A \\
\delta\Psi^A&=\Omega^{AB}\gamma^{\mu}\epsilon D_{\mu}X_B+\delta_3\Psi^A \\
\delta A_{\mu}&=\frac{1}{k_1}[\Omega_{AB}\tilde{\epsilon}\gamma_{\mu}\Psi^AX^B+\Omega^{AB}X_B\tilde{\Psi}_A\gamma_{\mu}\epsilon] \\
\delta \hat{A}_{\mu}&=\frac{1}{k_2}[\Omega_{AB}X^B\tilde{\epsilon}\gamma_{\mu}\Psi^A+\Omega^{AB}\tilde{\Psi}_A\gamma_{\mu}\epsilon X_B]~,
\end{split}
\end{equation}
where the variation $\delta_3$ will be determined in the following.

$\bullet$ The variation  of $\cl_{CS}$ cancels against the variation of the matter fields in $\cl_{kin}$. 

$\bullet$ The variation of the gauge fields in the spinor kinetic term in $\cl_{kin}$ cancels against the variation of the bosonic fields in $\cl_4$, iff:
\eq{
\spl{
&\frac{2}{k_1}+2\alpha_1-ia_1+2i\bar{a}_4=0~,~-\frac{2}{k_2}-2\bar{\alpha}_1-ia_2+2i a_4=0~,\\
&\frac{1}{k_1}+2\alpha_1+2i\bar{a}_4+\alpha_{2,1}=0~,~-\frac{1}{k_2}-2\bar{\alpha}_1+2ia_4-\alpha_{2,2}=0~,\\
&\frac{1}{k_2}+2\alpha_1+\alpha_{2,2}=0~,~\frac{1}{k_1}+2\bar{\alpha}_1+\alpha_{2,1}=0~,\\
&2\alpha_1-2i\bar{a}_3+2\alpha_{3,2}=0~,~2\bar{\alpha}_1+2ia_3+2\alpha_{3,1}=0~,\\
&2\alpha_1-2i\bar{a}_3+2\alpha_{3,1}=0~,~2\bar{\alpha}_1+2ia_3+2\alpha_{3,2}=0~,
}
}
or:
\eq{
\label{r1}
\spl{
a_1&= -2i(\frac{1}{k_1}+\bar{\alpha}_1)~,~a_2=2i(\frac{1}{k_2}+\alpha_1)~,\\
a_3&=-\bar{a}_3-i(\alpha_1-\bar{\alpha}_1)~,~a_4=\bar{a}_4=i(\alpha_1-\bar{\alpha}_1)~,\\
\alpha_{2,1}&=-\frac{1}{k_1}-2\bar{\alpha}_1~,~\alpha_{2,2}=-\frac{1}{k_2}-2\alpha_1~,\\
\alpha_{3,1}&= \alpha_{3,2}=i\bar{a}_3-\alpha_1=-ia_3-\bar{\alpha}_1~,
}
}
where all parameters are expressed in terms of $k_1$,$k_2$,$\alpha_1$,$\bar{\alpha}_1$ and $\bar{a}_3$. In the following we will set $\alpha_3\equiv\alpha_{3,1}= \alpha_{3,2}$, and use $a_4$ instead of $\bar{a}_4$.

$\bullet$ The variation of the gauge fields in the boson kinetic terms in $\cl_{kin}$, together with the variation of the fermion fields in $\cl_4$ without $\delta_3\Psi$, cancel against the $\delta_3\Psi$ variation of the fermion kinetic terms in $\cl_{kin}$, iff:
\eq{
\spl{
\delta_3\Psi_A&=\{\Omega_{AB}(\alpha_{2,2}X^CX_CX^B-\alpha_{2,1}X^BX_CX^C)-2\alpha_3\Omega_{BC}X^BX_AX^C\}\epsilon~,\\
\delta_3\Psi^A&=\{\Omega^{AB}(-\alpha_{2,1}X_CX^CX_B+\alpha_{2,2}X_BX^CX_C)+2\alpha_3\Omega^{BC}X_BX^AX_C\}\epsilon~.
}
}
\vfill\break

$\bullet$ The $\delta_3\Psi$ variation of $\cl_4$ cancels against the variation of $\cl_6$, iff:
\eq{
\spl{
&2i\alpha_1\alpha_{2,1}-i\alpha^2_{2,1}-a_1\alpha_{2,1}-2a_4\alpha_{2,1}-i\alpha_{4,2}+P = 0~, \\
&2i\bar{\alpha}_1\alpha_{2,2}-i\alpha^2_{2,2}+a_2\alpha_{2,2}+2a_4\alpha_{2,2}-i\alpha_{4,1}+\bar{P} = 0~, \\
&-2i\bar{\alpha}_1\alpha_{2,1}-a_2\alpha_{2,1}-2a_4\alpha_{2,1}+2i\alpha_{4,4}-\bar{P} = 0~, \\
&-2i\alpha_1\alpha_{2,2}+a_1\alpha_{2,2}+2a_4\alpha_{2,2}+2i\alpha_{4,4}-P = 0~, \\
&2i\alpha_1\alpha_{2,1}-2i\alpha_3\alpha_{2,1}+2\bar{a}_3\alpha_{2,1}-i\bar{m}+P=0~, \\
&2i\bar{\alpha}_1\alpha_{2,2}-2i\alpha_3\alpha_{2,2}-2a_3\alpha_{2,2}-im+\bar{P}=0~, \\
&-2i\alpha_1\alpha_{2,2}+2i\alpha_3\alpha_{2,2}-2\bar{a}_3\alpha_{2,2}+im-P=0~, \\
&-2i\bar{\alpha}_1\alpha_{2,1}+2i\alpha_3\alpha_{2,1}+2a_3\alpha_{2,1}+i\bar{m}-\bar{P}=0~, \\
&-2i\alpha_1\alpha_{2,1}-a_2\alpha_{2,1}+2i\alpha_{4,4}-P=0~, \\
&-2i\bar{\alpha}_1\alpha_{2,2}+a_1\alpha_{2,2}+2i\alpha_{4,4}-\bar{P}=0~, \\
&2i\alpha_1\alpha_{2,2}-i\alpha^2_{2,2}+a_2\alpha_{2,2}-i\alpha_{4,1}+P=0~, \\
&2i\bar{\alpha}_1\alpha_{2,1}-i\alpha^2_{2,1}-a_1\alpha_{2,1}-i\alpha_{4,2}+\bar{P}=0~, \\
&4i\alpha_{2,2}\alpha_3+im-P=0~,~4i\alpha_{2,1}\alpha_3+i\bar{m}-\bar{P}=0~, \\
&4i\alpha_{2,1}\alpha_3+i\bar{m}-P=0~,~4i\alpha_{2,2}\alpha_3+im-\bar{P}=0~, \\
&2i\alpha_{2,1}\alpha_{2,2}+2i\alpha_{4,4}-P=0~,~2i\alpha_{2,1}\alpha_{2,2}+2i\alpha_{4,4}-\bar{P}=0~, \\
&4i\alpha^2_3-in-P=0~,~4i\alpha^2_3-in-\bar{P}=0~, \\
&4i\alpha_{4,3}-P=0~,~4i\alpha_{4,3}-\bar{P}=0~,
}
}
and:
\begin{equation}
\begin{split}
&2a_1\alpha_3+4a_4\alpha_3+4\bar{a}_3\alpha_3+i\bar{m}+in=0~, \\
&2a_2\alpha_3-im+2a_1\alpha_3+4a_4\alpha_3+i\bar{m}=0~, \\
&-2a_2\alpha_3-4a_4\alpha_3-4a_3\alpha_3+im+in=0~, \\
&2a_1\alpha_3+i\bar{m}+2a_2\alpha_3+4a_4\alpha_3-im=0~, 
\end{split}
\end{equation}
where:
\begin{equation}
\begin{split}
P=-4i\alpha_1\alpha_3+2a_1\alpha_3+4a_4\alpha_3+i\bar{m}~, \\
\bar{P}=-4i\bar{\alpha}_1\alpha_3-2a_2\alpha_3-4a_4\alpha_3+im~,
\end{split}
\end{equation}
and we made use of the identities: 
\eq{
\varepsilon_{ABCD}=\Omega_{AB}\Omega_{CD}-\Omega_{AC}\Omega_{BD}+\Omega_{AD}\Omega_{BC}~;~~~
\varepsilon_{ABCD}\Omega^{EF}=24\Omega_{[AB}\delta^E_C\delta^F_{D]}
~.}
\vfill\break
After some further manipulation of these equations, taking (\ref{r1}) into account, we find:
\begin{equation}
\label{r2}
\begin{split}
a_1&=-2i(\frac{1}{k_1}+\bar{\alpha}_1),~~a_2=2i(\frac{1}{k_2}+\alpha_1)~,\\
a_3&=-\bar{a}_3-i(\alpha_1-\bar{\alpha}_1),~~a_4=i(\alpha_1-\bar{\alpha}_1)~,\\
\alpha_{2,1}&=-\frac{1}{k_1}-2\bar{\alpha}_1,~~\alpha_{2,2}=-\frac{1}{k_2}-2\alpha_1~,\\
\alpha_3&=i\bar{a}_3-\alpha_1=-ia_3-\bar{\alpha}_1~,\\
\alpha_{4,1}&=-3\alpha^2_{2,2}+4\alpha_{2,2}\alpha_3+m~,\\
\alpha_{4,2}&=-3\alpha^2_{2,1}+4\alpha_{2,2}\alpha_3+m~,\\
\alpha_{4,3}&=\alpha_{2,2}\alpha_3+\frac{m}{4}~,\\
\alpha_{4,4}&=-\alpha_{2,1}\alpha_{2,2}+2\alpha_{2,2}\alpha_3+\frac{m}{2}~,\\
\bar{m}&=4(\alpha_{2,2}-\alpha_{2,1})\alpha_3+m~\\
n&=4(\alpha_3-\alpha_{2,2})\alpha_3-m~.
\end{split}
\end{equation}
$\bullet$ Let us also mention that the requirement that the total Lagrangian should be real (which we do not need to impose in the present paper) would imply the following additional conditions on the parameters:
\begin{equation}
\begin{split}
&\alpha_{2,1},\alpha_{2,2},\alpha_3,\alpha_{4,1},\alpha_{4,2},\alpha_{4,3},\alpha_{4,4},m,\bar{m},n~\text{are real}~,\\
&a_1, a_2~\text{are imaginary}~,\\
&(a_3-i\bar{\alpha}_1)^\ast=\bar{a}_3+i\alpha_1~,~(a_4+i\bar{\alpha}_1)^\ast=-i\alpha_1~,~(i\bar{\alpha}_1)^\ast=a_4-i\alpha_1~.
\end{split}
\end{equation}
When we combine the above reality conditions with (\ref{r2}), we see that $\alpha_1$ and $\bar{\alpha}_1$ are real and $a_4$, $a_3$, $\bar{a}_3$ are imaginary.

From the above it follows that the on-shell theory has four independent parameters besides the CS levels $k_1$, $k_2$. They can be chosen to be $\alpha_1$, $\bar{\alpha}_1$, $\bar{a}_3$ and $m$.

\subsection{Conformal supersymmetry}\label{sec:conf}

Provided (\ref{r2}) holds, the action possesses an additional conformal supersymmetry. 
To show this, we follow \cite{Ooguri:2008dk} and replace the parameter $\epsilon$ of the Poincar\'e supersymmetry by $x_{\mu} \gamma^{\mu}\eta$, while adding to the spinor variations the terms:
\begin{equation}
\begin{split}
\delta^\prime\Psi_A&=\Omega_{AB}X^B\eta~, \\
\delta^\prime\Psi^A&=\Omega^{AB}X_B\eta~.
\end{split}
\end{equation}
Most terms in the Lagrangian are then invariant by virtue of the Poincar\'e supersymmetry. The term coming from the derivative acting on $x$ of $x_{\mu} \gamma^{\mu}\eta$ in $\delta_3 \Psi$ of the fermion kinetic Lagrangian cancels with $\delta^\prime\Psi$ of $\cl_4$, if (\ref{r2}) holds. Finally terms generated by the remaining variations of the fermions in the fermionic kinetic terms cancel against the boson transformations in the bosonic kinetic Lagrangian and the variations of the CS terms.

\section{Trivial $G$-structures in 3d}

The existence of a nowhere-vanishing (commuting) spinor $\epsilon$ on a Riemannian three-manifold 
implies the existence of a trivial $G$-structure, i.e. the trivialization 
of the tangent bundle. In this section we will explore in detail the implications of this trivialization.

Since $\epsilon$ is assumed nowhere-vanishing we can take it to be normalized:
\eq{\epsilon^\dagger\epsilon=\tilde{\epsilon}\epsilon^c=-\tilde{\epsilon^c}\epsilon=1
~,}
where we used the formulas in Appendix \ref{app:sc}. On the other hand,
\eq{\tilde{\epsilon}\epsilon=0~,}
due to the antisymmetry of the charge conjugation matrix, cf. (\ref{a4}). 
furthermore we can define the following $\epsilon$-bilinear one-forms:\footnote{Since we are assuming the existence of a Riemannian 
metric on our manifold, we can convert vectors to one-forms and vice-versa.}
\eq{U_{\mu}\equiv
\epsilon^\dagger\g_{\mu}\epsilon=-\tilde{\epsilon^c}\g_{\mu}\epsilon=-\tilde{\epsilon}\g_{\mu}\epsilon^c
~,}
where we took (\ref{a6}) into account, and:
\eq{V_{\mu}\equiv
\tilde{\epsilon}\g_{\mu}\epsilon
~.}
It can be seen that $U$ is real whereas $V$ is complex:
\eq{
\bar{V}_{\mu}=-\tilde{\epsilon^c}\g_{\mu}\epsilon^c=\epsilon^{\dagger}\g_{\mu}\epsilon^c
~.}
The Fierz identities can be conveniently written in terms of the  
bilinears above:
\eq{\spl{\label{fierz}
\epsilon\tilde{\epsilon^c}=-\frac12(\bbone +U^{\mu}\gamma_{\mu})~&;~~~
\epsilon^c\tilde{\epsilon}=\frac12(\bbone -U^{\mu}\gamma_{\mu})\\
\epsilon\tilde{\epsilon}=\frac12 V^{\mu}\gamma_{\mu}~&;~~~
\epsilon^c\tilde{\epsilon^c}=-\frac12 \bar{V}^{\mu}\gamma_{\mu}
~.}}
Using the above, the following relations can be shown:
\eq{\label{ortho}U^2=\Re V^2=\Im V^2=1~;~~~U\cdot \Re V=U\cdot \Im V=\Re V\cdot \Im V=0~,}
where we have defined $A^2\equiv A^{\mu}A_{\mu}$, $A\cdot B\equiv A^{\mu}B_{\mu}$ and $V=\Re V+i\Im V$. In other words the triplet ($U$, $\Re V$, $\Im V$) is a globally-defined orthonormal frame thus trivializing the (co)tangent bundle of 
the manifold.

Let us also mention the following useful identities which can similarly be shown 
by fierzing:
\eq{\label{gdec}\spl{\g_{\mu}\epsilon&=U_{\mu}\epsilon+V_{\mu}\epsilon^c\\
\g_{\mu}\epsilon^c&=\bar{V}_{\mu}\epsilon-U_{\mu}\epsilon^c
~.}}
From these we also obtain:
\eq{\spl{\label{sd}
U^{\mu}\g_{\mu}\epsilon=\epsilon~;~&~~U^{\mu}\g_{\mu}\epsilon^c=-\epsilon^c\\
\frac12 V^{\mu}\g_{\mu}\epsilon^c=\epsilon~;~&~~\frac12\bar{V}^{\mu}\g_{\mu}\epsilon=\epsilon^c\\
V^{\mu}\g_{\mu}\epsilon=&\bar{V}^{\mu}\g_{\mu}\epsilon^c=0
~.
}}

\subsubsection*{Spinor and tensor decomposition}

Spinors on the manifold can be expanded on the basis of $\epsilon$, $\epsilon^c$. 
Explicitly, for any spinor $\lambda$ we have:
\eq{\lambda=\lambda_+\epsilon+\lambda_-\epsilon^c~,}
where $\lambda_{\pm}$ are scalar coefficients given by:
\eq{\lambda_+=\tilde{\lambda}\epsilon^c~;~~~\lambda_-=\tilde{\epsilon}\lambda~.}
The notation is motivated by the fact that 
we may define a chirality operator:
\eq{\gamma\equiv U^{\mu}\g_{\mu}~,}
which indeed squares to one as follows from (\ref{ortho}). Moreover 
$\epsilon$, $\epsilon^c$ are chiral, antichiral respectively with respect ot $\gamma$, as can be seen from (\ref{sd}).

Forms and tensors can be decomposed using the orthonormal frame provided by $(U,V)$. For example any one-form $A$ can be decomposed as follows:
\eq{A=A_{\perp}U+A_+V+A_-\bar{V}~,}
where $A_{\perp}$, $A_{\pm}$ are scalar coefficients given by:
\eq{A_{\perp}=U\cdot A~;~~~A_+=\frac12\bar{V}\cdot A~;~~~A_-=\frac12 V\cdot A~.}
The notation is motivated by the fact that one-forms can be decomposed into the 
subspaces parallel and orthogonal to $U$, which we may call the vertical and horizontal subspaces respectively. The horizontal subspace can then be further 
decomposed into directions parallel and orthogonal to $V$ (equivalently: orthogonal and parallel to $\bar{V}$), which we may consider as the holomorphic and antiholomorphic directions respectively.

\vfill\break

\section{Lorentz Gauge}

In this section we give the details of the integration over the ghost complex. 
First note that in (\ref{r7}) the integration over $b$ and $b_0$ can be performed independently:
\begin{equation}
\begin{split}
&\int\prod_xdb(x)db_0\exp^{i\int dx^3 \mathrm{tr} \{b(\partial^\mu A_\mu+b_0)\}}\\
=&\int\prod_xdb^\prime(x)db_0\exp^{i\int dx^3 \mathrm{tr} \{b^\prime(x)(\partial^\mu A_\mu+b_0)\}}\int db^\prime\exp^{i\int dx^3 \mathrm{tr} \{b^\prime(\partial^\mu A_\mu+b_0)\}}\\
=&\int\prod_xdb^\prime(x)db_0\exp^{i\int dx^3 \mathrm{tr} \{b^\prime(x)(\partial^\mu A_\mu+b_0)\}}\int db^\prime\exp^{i\int dx^3 \mathrm{tr} \{b^\prime b_0\}}\\
=&\frac{1}{\text{Vol}}\int\prod_xdb^\prime(x)db_0\exp^{i\int dx^3 \mathrm{tr} \{b^\prime(x)(\partial^\mu A_\mu+b_0)\}}\delta(b_0)\\
=&\int\prod_xdb^\prime(x)db_0\exp^{i\int dx^3 \mathrm{tr} \{b^\prime(x)(\partial^\mu A_\mu+b_0)\}}\int db^\prime\exp^{i\int dx^3 \mathrm{tr} \{b^\prime b_0\}}\\
=&\frac{1}{\text{Vol}}\int\prod_xdb^\prime(x)\exp^{i\int dx^3 \mathrm{tr} \{b^\prime(x)\partial^\mu A_\mu\}}~,
\end{split}
\end{equation}
where \text{Vol} denotes the volume of $T^3$, and we decompose $b(x) = b^\prime(x)+b^\prime$; $b^\prime$ is a constant field: it is the zero mode of $b(x)$. The remaining integration over $b^\prime(x)$ imposes the Lorentz gauge condition.

Next we integrate over $a_0$, then $\bar{a}_0$:
\begin{equation}
\begin{split}
&\int d\bar{a}_0 da_0\exp^{i\int dx^3 \mathrm{tr} \{-(a_0-\frac{i}{2}\{C,C\})\bar{a}_0\}}\\
=&\frac{1}{\text{Vol}}\int d\bar{a}_0\exp^{i\int dx^3 \mathrm{tr} \{\frac{i}{2}\{C,C\}\bar{a}_0\}}\delta(\bar{a}_0)\\
=&\frac{1}{\text{Vol}}~.
\end{split}
\end{equation}

\vfill\break

The remaining integrations read:
\begin{equation}
\begin{split}
&\int\prod_xd\bar{C}(x)dC(x)dC_0d\bar{C}_0\exp^{i\int dx^3 \mathrm{tr} \{-\bar{C}(\partial^\mu D_\mu C+\partial^\mu\delta_QA_\mu+C_0)+C\bar{C}_0\}}\\
=&\int\prod_xdC(x)dC_0d\bar{C}_0d\bar{C}^\prime(x)\exp^{i\int dx^3 \mathrm{tr} \{-\bar{C}^\prime(x)(\partial^\mu D_\mu C+\partial^\mu\delta_QA_\mu+C_0)+C\bar{C}_0\}}\\
&\times\int d\bar{C}^\prime\exp^{i\int dx^3 \mathrm{tr} \{-\bar{C}^\prime(\partial^\mu D_\mu C+\partial^\mu\delta_QA_\mu+C_0)\}}\\
=&\int\prod_xdC(x)dC_0d\bar{C}_0\int d\bar{C}^\prime(x)\exp^{i\int dx^3 \mathrm{tr} \{-\bar{C}^\prime(x)(\partial^\mu D_\mu C+\partial^\mu\delta_QA_\mu+C_0)+C\bar{C}_0\}}\\
&\times\int d\bar{C}^\prime\exp^{i\int dx^3 \mathrm{tr} \{-\bar{C}^\prime C_0\}}\\
=&\int\prod_xdC(x)dC_0d\bar{C}_0\int \bar{C}^\prime(x)\exp^{i\int dx^3 \mathrm{tr} \{-\bar{C}^\prime(x)(\partial^\mu D_\mu C+\partial^\mu\delta_QA_\mu+C_0)+C\bar{C}_0\}}\\
&\times\text{Vol}\delta(C_0)\\
=&\text{Vol}\int\prod_xdC(x)d\bar{C}_0\int d\bar{C}^\prime(x)\exp^{i\int dx^3 \mathrm{tr} \{-\bar{C}^\prime(x)(\partial^\mu D_\mu C+\partial^\mu\delta_QA_\mu)+C\bar{C}_0\}}\\
=&\text{Vol}\int\prod_xdC^\prime(x)d\bar{C}_0\int d\bar{C}^\prime(x)\exp^{i\int dx^3 \mathrm{tr} \{-\bar{C}^\prime(x)(\partial^\mu D_\mu C^\prime(x)+\partial^\mu\delta_QA_\mu)+C^\prime(x)\bar{C}_0\}}\\
&\times\int dC^\prime\exp^{i\int dx^3 \mathrm{tr} \{-i\bar{C}^\prime(x)[\partial^\mu A_\mu,C^\prime]+C^\prime\bar{C}_0\}}~.
\end{split}
\end{equation}
Note that the expression above is multiplied by an overall factor $\delta(\partial\cdot A)$, therefore we can set $\partial\cdot A$ to zero and integrate over $C^\prime$:
\begin{equation}
\begin{split}
&\text{Vol}^2\int\prod_xdC^\prime(x)d\bar{C}^\prime(x)d\bar{C}_0\exp^{i\int dx^3 \mathrm{tr} \{-\bar{C}^\prime(x)(\partial^\mu D_\mu C^\prime(x)+\partial^\mu\delta_QA_\mu)+C^\prime(x)\bar{C}_0\}}\\
&\times\delta(\bar{C}_0)\\
=&\text{Vol}^2\int\prod_xdC^\prime(x)d\bar{C}^\prime(x)\exp^{i\int dx^3 \mathrm{tr} \{-\bar{C}^\prime(x)(\partial^\mu D_\mu C^\prime(x)+\partial^\mu\delta_QA_\mu)\}}~.
\end{split}
\end{equation}
Absorbing $\delta_QA_\mu$ into $C^\prime(x)$, 
restricting to the saddle point $A=0$ and integrating 
over $C'$ and $\bar{C}'$, 
the last line gives $\det\DAlambert$.

To see that  $\delta_QA_\mu$ can indeed be absorbed into $C^\prime(x)$ it suffices to show  that there is a $C^{\prime\prime}$ such that:
\eq{  \partial^\mu D_\mu C^\prime(x)+ \partial^\mu \delta_QA_\mu
=\partial^\mu D_\mu C^{\prime\prime}
~.}

Equivalently in form notation:
\eq{\label{1} d^{\dagger} (\delta_QA+ d_A \delta C)=0 ~,}
where we have set $\delta C:=C^{\prime}-C^{\prime\prime}$, $d_A :=d+i[A,]$. The  Hodge decompositions of $\delta C$,  $\delta_QA$ are as follows:
\eq{\label{e1}\delta C= \delta C_{(h)}+d^{\dagger}\delta C_{(1)}~;~~~\delta_QA=\delta_QA_{(h)}+d\delta_QA_{(0)}+
d^{\dagger}\delta_QA_{(2)}~,
}
where the numerical subscripts indicate the rank of the corresponding form and 
$\delta C_{(h)}$, $\delta_QA_{(h)}$ are harmonic zero-, one-forms respectively; in particular $\delta C_{(h)}$ is constant. Similarly for the gauge field  we expand:
\eq{\label{e2}A=
d^{\dagger} A_{(2)}+ A_{(h)}~.}
The fact  that there is no exact  piece in the decomposition above is due to the Lorentz gauge, $d^{\dagger}A=0$. 
Furthermore equation (\ref{1}) is equivalent to the statement that there exist 
a two-form $u$ and a harmonic one-form $w_h$ such that:
\eq{\label{2}d_A \delta C+\delta_QA=d^{\dagger}u+w_{(h)}~.}
On the other hand, taking the expansions (\ref{e1}),(\ref{e2}) into account, the left-hand side of (\ref{1}) reads:
\eq{d( \delta C+\delta_QA_{(0)})
+id^{\dagger}([A_{(2)}, \delta C]+[A_{(h)}, \delta C_{(1)}])
+i[A_{(h)}, \delta C_{(h)}]
~.}
It follows that (\ref{2}), is solved for:
\eq{ \delta C=-\delta_QA_{(0)}~;~~~
u=i([A_{(2)}, \delta C]+[A_{(h)}, \delta C_{(1)}])~;~~~
w_{(h)}=i[A_{(h)}, \delta C_{(h)}]
~.}

{}

\end{document}